\documentclass[aps,prd,preprint,nofootinbib,superscriptaddress,11pt]{revtex4-2}

\usepackage[T1]{fontenc}
\usepackage[utf8]{inputenc}
\usepackage{lmodern}

\usepackage{amsmath}
\usepackage{amssymb}
\usepackage{amsfonts}
\usepackage{amsthm}
\usepackage{mathtools}
\usepackage{mathrsfs}
\usepackage{bm}

\usepackage{graphicx}
\usepackage{xcolor}
\usepackage{booktabs}
\usepackage{multirow}

\usepackage[
    colorlinks=true,
    linkcolor=blue,
    citecolor=blue,
    urlcolor=blue
]{hyperref}

\newcommand{\scri}{\mathscr{I}}
\newcommand{\dd}{\mathrm{d}}
\newcommand{\ii}{\mathrm{i}}
\newcommand{\ee}{\mathrm{e}}
\newcommand{\p}{\partial}

\newcommand{\ethb}{\bar{\eth}}
\newcommand{\Order}{\mathcal{O}}
\newcommand{\PV}{\operatorname{PV}}
\newcommand{\ep}{\varepsilon}
\renewcommand{\eth}{\text{\dh}}

\newcommand{\Amap}{\mathcal{A}}

\begin{document}

\title{Quantum graviton scattering with definite helicities
in the null surface formulation. III:\\
Fourth-order recursion and uniform radial ultraviolet convergence}

\author{C.~N.~Kozameh}
\email{carlos.kozameh@unc.edu.ar}
\affiliation{FaMAF, Universidad Nacional de C\'ordoba, 
5000 C\'ordoba, Argentina}

\author{G.~O.~Depaola}
\affiliation{FaMAF, Universidad Nacional de C\'ordoba, 
5000 C\'ordoba, Argentina}

\date{\today}

\begin{abstract}
We extend the helicity-resolved null-surface formulation (NSF) of graviton
scattering through fourth order. The resulting outgoing correction identifies
the mixed contributions $\mathcal M_{24}+\mathcal M_{42}$ which, together with
the previously obtained $\mathcal M_{33}$, complete the two-vertex one-loop
four-graviton sector at the same perturbative order. We give the general quantum
matching relation in recursive form, retaining the operator-valued cut phase
$U_\omega$. For comparison with conventional perturbative graviton amplitudes,
we subsequently adopt the Minkowskian prescription of Part II: $U_\omega=1$,
with the metric, null trajectories, and extraction cuts kept classical. The
background phase and nonlinear NSF interaction sources remain. Combining
radiative normalization, canonical contractions, recursive principal-value
denominators, and loop measures yields the uniform radial estimate
$\int^\infty dK/K^{2q-2}$ for connected flat-cone processes with arbitrary
incoming multiplicity $p$, outgoing multiplicity $q\geq2$, and loop number
$L\geq1$; the exponent does not depend on $L$. Both one-loop sectors
therefore have radial behavior $\int^\infty dK/K^2$. This convergence result applies to common hard-momentum
scaling with fixed generic ratios and directions; nonuniform ultraviolet
regions and infrared or angular singularities are not analyzed.
\end{abstract}

\maketitle

%=====================================================================
\section{Introduction}
\label{sec:introduction}
%=====================================================================
The null-surface formulation (NSF) provides a framework for studying
quantum gravity in terms of radiative data at null infinity. Its basic
geometrical variable is the cut function $Z(x^a,\zeta,\bar\zeta)$, whose
level surfaces are null and whose angular derivatives determine the
conformal geometry. The Bondi shear supplies the radiative degrees of
freedom from which the spacetime fields are reconstructed. Quantizing the
asymptotic shear gives incoming and outgoing graviton operators at
$\scri^-$ and $\scri^+$, respectively.

Asymptotic quantization fixes the algebra of the free radiative modes;
the NSF field equations provide the dynamics relating the incoming and
outgoing operators. In this series the two asymptotic reconstructions are
matched to describe the same spacetime, giving a perturbative scattering
map in terms of the incoming modes \cite{KDI2026,KDII2026}. The first two
papers develop this construction through third order and apply it to
helicity-resolved scattering.

The operator map can also be organized through the
Baker--Campbell--Hausdorff (BCH) expansion of a perturbatively reconstructed
unitary transformation \cite{KozamehBCH2026,Hall2015,PeskinSchroeder}.
Its nested commutators are evaluated using canonical contractions: terms
with no contraction between the commutator blocks cancel. This organizes
the resulting contributions as diagrams linked to the transformed operator
insertion. Connected scattering amplitudes are then obtained by selecting
contributions connecting the external particles; the BCH linkage by itself
does not eliminate spectator terms in multiparticle matrix elements.

In this work we extend the NSF hierarchy and outgoing-operator construction to fourth
order. The third-order calculation of Part II identified the cubic--cubic
contribution $\mathcal M_{33}$ to the one-loop four-graviton sector. At
the same perturbative order, the two-vertex sector also contains the mixed
contributions $\mathcal M_{24}+\mathcal M_{42}$. Their identification
requires the fourth-order outgoing correction. We determine their topology
and uniform radial ultraviolet behavior, and then extend the power counting
to connected flat-cone processes with arbitrary incoming multiplicity $p$,
outgoing multiplicity $q\geq2$, and loop number $L\geq1$.

The construction is presented in two stages. We first develop the
fourth-order NSF field equations and state the general quantum matching
result, retaining the operator-valued cut phase $U_\omega$ and the
recursive outgoing correction. The technical expansion of $U_\omega$
and the intermediate matching calculations are collected in Appendix~\ref{sec:nsf_yang_feldman}.
We then adopt the Minkowskian comparison prescription of Part II:
$U_\omega=1$, with the metric, null trajectories, and extraction cuts
kept classical. The background phase $e^{-i\omega Z_0}$ and the nonlinear
NSF interaction sources remain. The mixed loop sector and the ultraviolet
bound are evaluated in this comparison sector, rather than for the full
operator-valued cut construction.

The complete momentum-space kernel of
$\delta a_{4,\lambda}^{\mathrm{out}}$ is not needed here. Its operator
degree and the homogeneity of the fourth-order geometrical source suffice to
characterize the mixed sector
\begin{equation}
\mathcal{M}_{24}
\equiv
\mathcal{M}^{(24)}+\mathcal{M}^{(42)},
\end{equation}
to identify its one-loop topology, and to determine its uniform radial
ultraviolet behavior. The connected four-point sector at absolute
perturbative order six consequently has the formal decomposition
\begin{equation}
\boxed{
\mathcal{M}^{(6)}_{2\to2}
=
\mathcal{M}_{33}+\mathcal{M}_{24}.
}
\label{eq:intro_complete_order_six}
\end{equation}
This equation identifies the two perturbative sectors contributing at that
order; it does not imply that the complete momentum-space kernel of
$\mathcal{M}_{24}$ is evaluated explicitly.

Section~\ref{sec:fourth_order_hierarchy} gives the fourth-order field equations. Section~\ref{sec:construction_guide} states the quantum matching result, its recursion, and the subsequent comparison prescription, and Sec.~\ref{sec:mixed_four_point} identifies the
mixed one-loop sector. Section~\ref{sec:UV_general_order} gives the
uniform radial ultraviolet bound. The detailed matching algebra and
lower-order checks are collected in Appendix~\ref{sec:nsf_yang_feldman};
Appendix~\ref{app:scalar_yang_feldman} explains the scalar analogy.

\section{Fourth-order NSF hierarchy}
\label{sec:fourth_order_hierarchy}
%=====================================================================

We first give the fourth-order field equations. Their sources depend on the lower-order NSF fields constructed in Parts I and II.
Following the notation of the second paper, define
\begin{align}
P_n
&\equiv
\sum_{i+j=n}
\p_r\Lambda_i\p_r\bar\Lambda_j,
&
A_n
&\equiv
\sum_{i+j=n}
\p_r^2\Lambda_i\p_r^2\bar\Lambda_j.
\label{eq:Pn_An_III}
\end{align}
In particular,
\begin{align}
P_2
&=
\p_r\Lambda_1\p_r\bar\Lambda_1,
&
A_2
&=
\p_r^2\Lambda_1\p_r^2\bar\Lambda_1,
\label{eq:P2_A2_III}
\end{align}
and the second-order conformal factor is
\begin{equation}
\Omega_2(r)
=
\frac18 P_2(r)
+
\frac18
\int_r^\infty \dd r'
\int_{r'}^\infty \dd r''\,
A_2(r'').
\label{eq:Omega2_integrated_III}
\end{equation}
At fourth order, the exact expansion of the rational radial equation gives
\begin{align}
\begin{aligned}
2\p_r^2\Omega_4
=
\frac14\left(A_4+\p_r^2P_4\right)
&+
\frac14P_2\left(A_2+\p_r^2P_2\right)
+
\frac38(\p_rP_2)^2
\\
&+
\frac14\left(A_2+\p_r^2P_2\right)\Omega_2,
\end{aligned}
\label{eq:Omega4_III}
\end{align}
where
\begin{align}
P_4
={}&
\p_r\Lambda_1\p_r\bar\Lambda_3
+
\p_r\Lambda_2\p_r\bar\Lambda_2
+
\p_r\Lambda_3\p_r\bar\Lambda_1,
\label{eq:P4_III}
\\
A_4
={}&
\p_r^2\Lambda_1\p_r^2\bar\Lambda_3
+
\p_r^2\Lambda_2\p_r^2\bar\Lambda_2
+
\p_r^2\Lambda_3\p_r^2\bar\Lambda_1.
\label{eq:A4_III}
\end{align}
The last term in Eq.~\eqref{eq:Omega4_III} is the first contribution generated
by the product $R_{rr}^{(2)}\Omega_2$.

With the standard boundary conditions at radial infinity, the same equation
may be written in the integrated form
\begin{equation}
\boxed{
\begin{aligned}
\Omega_4(r)
={}&
\frac18 P_4(r)
+
\int_r^\infty \dd r'
\int_{r'}^\infty \dd r''\,
\Bigg[
\frac18 A_4
+
\frac18P_2\left(A_2+\p_r^2P_2\right)
\\
&\hspace{33mm}
+
\frac{3}{16}(\p_rP_2)^2
+
\frac18\left(A_2+\p_r^2P_2\right)\Omega_2
\Bigg](r'').
\end{aligned}
}
\label{eq:Omega4_integrated_III}
\end{equation}
The integrated form displays the local fourth-order term $P_4/8$.
Additional local polynomials, including the $P_2^2$ contribution, arise when
the radial derivatives in the lower-order rational terms are integrated by
parts.
These local terms determine the leading power of the large internal momentum
in the ultraviolet analysis. Indeed, under a uniform routing,
$\p_r=\mathcal O(K)$ whereas each radial integration contributes
$\mathcal O(K^{-1})$. Higher radial derivatives in the source are therefore
accompanied by the integrations required to solve the radial equation, and
the overall radial degree remains homogeneous. The ultraviolet degree of
$\Omega_n$ can consequently be read from its local polynomial part in the
$\Lambda_j$ and lower-order $\Omega_j$, subject only to possible cancellations
among terms of the same degree.

The fourth-order cone source is
\begin{equation}
\boxed{
\mathcal{J}_4(x,z')
=
-
\left[
\int_s^\infty
\left(
\ethb(\eth\Omega^2)
+
g^{ab}\p_a\Lambda\p_b\bar\Lambda
\right)\dd s'
\right]_4,
}
\label{eq:I4_III}
\end{equation}
with $\Omega_4$ obtained from Eq.~\eqref{eq:Omega4_III}. In accordance with
the working hypothesis, every occurrence of $Z_2$, $\Lambda_2$, $Z_3$,
$\Lambda_3$, $\Omega_2$, $\Omega_3$, and the corresponding metric
coefficients is replaced by its retarded functional of the incoming free
data. This defines the single fourth-order source
\begin{equation}
    \widehat{\mathcal{J}}_4^{\mathrm{YF}}
    =
    \widehat{\mathcal{J}}_4
    [a^{\mathrm{in}},a^{\mathrm{in}\dagger}].
    \label{eq:I4_YF_incoming_III}
\end{equation}
The mapped future cone and the past cone use this same source. Their sum is
therefore
\begin{equation}
    \boxed{
    \widehat Z_{4,\mathrm{cone}}^{\mathrm{YF}}
    =
    2\ii\int \dd\Gamma_4\,
    \widehat{\mathcal{J}}_4^{\mathrm{YF}}
    \PV\frac{1}{D_4^{\mathrm{out}}}.
    }
    \label{eq:order4_common_PV_cone}
\end{equation}
This is the fourth-order use of the working prescription that was implicit in
the order-two and order-three constructions.

\section{Quantum matching and the comparison prescription}
\label{sec:construction_guide}
The purpose of extending the hierarchy is to identify the mixed one-loop
contribution missing from the third-order calculation. We use the
perturbative method of Parts I and II \cite{KDI2026,KDII2026}; the method
itself is not a new assumption at each order. The steps needed here are
summarized below. Appendix~\ref{sec:nsf_yang_feldman} gives the detailed
matching derivation and the checks through fourth order.

\subsection{Physical data and approximation}
The independent data are the free incoming Bondi shear modes. The NSF
fields reconstruct the interior geometry from these data, and matching
relates the outgoing shear to the incoming shear. The outgoing corrections
$\delta a_n^{\rm out}$ are therefore the quantities from which scattering
matrix elements are formed; the cut coefficients and cone sources are
intermediate quantities used to determine them.

For comparison with the perturbative graviton amplitudes of Weinberg and
collaborators, the scattering benchmark retains the Minkowski metric,
its null geodesics, and the background cuts used in radiative extraction.
This is the fixed-background comparison sector introduced in Part II.
It does not set the interaction-generated cut coefficients $Z_n$ to zero.
We distinguish the general matching construction from the flat-cone
vertices used in the ultraviolet proposition below. The latter exclude
explicit metric insertions and deformations of the background null
trajectories. The Poincar\'e reduction of the external data remains a
separate step, as in Parts I and II; the external four-momentum restriction
it uses is discussed in Ref.~\cite{KDIV2026}.

\subsection{Matching and recursive determination}
The Yang--Feldman prescription first determines local fields as retarded
functionals of the incoming data \cite{YangFeldman}. Both cone
representations then use this same local solution; their outer affine
integrations have opposite orientations. Their sum produces the
principal-value denominator in Eq.~\eqref{eq:PV_condition_III}.

The complete matching condition is
\begin{equation}
\Delta\widehat Z_\sigma+\Delta\widehat Z_\Sigma
+\widehat Z_{\rm cone}^{\rm YF}=0.
\label{eq:guide_matching}
\end{equation}
Here $\Delta\widehat Z_\sigma$ contains the difference of the radiative
cut data, $\Delta\widehat Z_\Sigma$ is the scalar News-flux contribution,
and $\widehat Z_{\rm cone}^{\rm YF}$ is the matched cone source.
The flux term belongs to the cut, rather than the cone source.
At order $n$, all source terms and recursive cut terms are fixed by lower
orders; the only new radiative unknown is $\delta a_n^{\rm out}$.
Fourier extraction and helicity projection then isolate that unknown.
The full momentum-convolution formula is
Eq.~\eqref{eq:delta_an_momentum_recursion_III}.

The ordered cut phase is organized by
\begin{equation}
\widehat Z_{[N]}=\sum_{j=1}^N\varepsilon^j\widehat Z_j,
\qquad
\widehat U_{\omega,[N]}=\exp[-i\omega\widehat Z_{[N]}].
\label{eq:guide_cut_phase}
\end{equation}
Its expansion supplies the lower-order products needed in matching.
No commutativity of different cut coefficients is assumed. The
noncommuting factor recursion and explicit lower-order checks are given
in Appendix~\ref{sec:nsf_yang_feldman}. Unitary realizations require the
self-adjoint domain assumptions stated there; their boundedness is not
used as a substitute for ultraviolet power counting.

The fourth-order source above supplies the cone term in the matching equation. The final fourth-order relation is stated below; its derivation is given in Appendix~\ref{sec:nsf_yang_feldman}.

The order-four matching equation in the incoming representation is
\begin{equation}
\boxed{
\widehat Z_{4,\mathrm{cut}}^{+}
[\delta a_4^{\mathrm{out}}]
+
\widehat{\mathscr Z}_{4}^{\mathrm{cut,rec}}
+
\widehat Z_{4,\Sigma}^{\mathrm{YF}}
=
-
\widehat Z_{4,\mathrm{cone}}^{\mathrm{YF}}.
}
\label{eq:order4_matching_III}
\end{equation}
The scalar flux term is thus grouped with the other cut contributions and is
not included in the cone source. The radiative helicity projection of
Eq.~\eqref{eq:order4_matching_III} formally determines
$\delta a_{4,\lambda}^{\mathrm{out}}$. In the present work we do not perform
the complete momentum-space evaluation of this operator. For the mixed
four-point topology and the ultraviolet estimate, only its quartic operator
content and the radial degree of the fourth-order geometrical kernel are
required.

%=====================================================================

\subsection{Recursive outgoing correction}
Let $[\cdot]_{F,\lambda}$ denote spatial Fourier extraction and helicity
projection, and let $\widehat{\mathcal U}_{\omega,r}$ be the coefficient
of $\varepsilon^r$ in the expansion of the cut-phase operator, with
$\widehat{\mathcal U}_{\omega,0}=1$. The general recursion is
\begin{equation}
\begin{aligned}
\delta a_{n,\lambda}^{\rm out}(\mathbf K)
={}&-\left[\widehat Z_{n,\rm cone}^{\rm YF}
+\widehat Z_{n,\Sigma}^{\rm YF}\right]_{F,\lambda}(\mathbf K)\\
&-\sum_{m=2}^{n-1}\int\frac{d^3q}{(2\pi)^3}
\delta a_{m,\lambda}^{\rm out}(\mathbf q)
\left[\widehat{\mathcal U}_{\omega_q,n-m}\right]_F(\mathbf K-\mathbf q).
\end{aligned}
\label{eq:main_general_recursion_v2}
\end{equation}
Here $\omega_q=|\mathbf q|$. The cut-phase coefficients and the scalar
flux term are fixed by lower orders. No commutativity of the cut
coefficients is assumed. At fourth order, the recursion takes the explicit form
\begin{equation}
\boxed{
\begin{aligned}
\delta a_{4,\lambda}^{\rm out}
={}&-
\left[
\widehat Z_{4,\rm cone}^{\rm YF}
+
\widehat Z_{4,\Sigma}^{\rm YF}
\right]_{F,\lambda}
+
\ii\omega\delta a_{3,\lambda}^{\rm out}\widehat Z_1
+
\ii\omega\delta a_{2,\lambda}^{\rm out}\widehat Z_2
\\
&+
\frac{\omega^2}{2}
\delta a_{2,\lambda}^{\rm out}\widehat Z_1^2,
\end{aligned}
}
\label{eq:order4_matching_from_exact_cut_III}
\end{equation}
The products are understood in the $x$ representation or, after Fourier
transformation, as momentum convolutions; their operator order is retained.
The cone source, scalar flux and recursive terms are determined by the incoming
data and lower-order equations. Appendix~\ref{sec:nsf_yang_feldman} gives
the ordered cut-phase expansion and the derivation of this expression.

\subsection{Minkowskian prescription for the scattering comparison}
The preceding equations describe the general matching construction.
For the loop comparison we now apply the simultaneous approximations
introduced in Part II \cite{KDII2026}:
\begin{equation}
\widehat U_\omega^-=1,\qquad \Sigma_2=0,\qquad
 g^{ab}_{\rm eval}=\eta^{ab},\qquad
 Z_{\rm eval}=Z_0,\qquad x_s^a=x_s^{(0)a}.
\label{eq:comparison_prescription_v2}
\end{equation}
The affine parameter and cone measure also retain their Minkowski values.
The extraction geometry is classical: the additional operator-valued cut
phase is omitted, while the background phase $e^{-i\omega Z_0}$ remains.
The nonlinear NSF sources are retained. Thus this prescription does not
set the interaction-generated coefficients $Z_n$ to zero.

As in Part II, the comparison omits the recursive ordered-cut and
nonlinear scalar-flux contributions, together with explicit metric and
trajectory corrections. The outgoing coefficient used in the following
flat-cone scattering and ultraviolet calculation is therefore
\begin{equation}
\delta a_{n,M}^{\rm out}
=-\left[\widehat Z_{n,\rm cone}^{\rm YF,flat}\right]_{F,\lambda}.
\label{eq:comparison_outgoing_v2}
\end{equation}
The subscript $M$ distinguishes this coefficient from the complete
operator. In the following flat-cone matrix elements it is understood
on each outgoing correction. The Poincar\'e reduction of external
kinematics is a further restriction, as in Parts I and II.

\section{The mixed four-point sector}
\label{sec:mixed_four_point}
%=====================================================================

The new contribution to connected $2\to2$ scattering is
\begin{align}
\mathcal{M}^{(24)}
={}&
\left\langle0\right|
\delta a_{2,\lambda'_1}^{\mathrm{out}}(p'_1)
\delta a_{4,\lambda'_2}^{\mathrm{out}}(p'_2)
 a_{\lambda_1}^{\mathrm{in}\dagger}(p_1)
 a_{\lambda_2}^{\mathrm{in}\dagger}(p_2)
\left|0\right\rangle_{\mathrm{conn}},
\label{eq:M24_definition_III}
\\
\mathcal{M}^{(42)}
={}&
\left\langle0\right|
\delta a_{4,\lambda'_1}^{\mathrm{out}}(p'_1)
\delta a_{2,\lambda'_2}^{\mathrm{out}}(p'_2)
 a_{\lambda_1}^{\mathrm{in}\dagger}(p_1)
 a_{\lambda_2}^{\mathrm{in}\dagger}(p_2)
\left|0\right\rangle_{\mathrm{conn}}.
\label{eq:M42_definition_III}
\end{align}
We define
\begin{equation}
\mathcal{M}_{24}
\equiv
\mathcal{M}^{(24)}+\mathcal{M}^{(42)}.
\label{eq:M24_symmetrized_III}
\end{equation}
Since the product contains two operators from $\delta a_2$, four from
$\delta a_4$, and two external creation operators, the connected Wick sector
contains eight radiative operators and is not excluded by operator parity.
After the two external contractions, the four remaining vertex-operator ends
form two internal lines. With two effective vertices, this gives
\begin{equation}
I_{24}=2,
\qquad
L_{24}=I_{24}-1=1,
\label{eq:M24_loop_topology_III}
\end{equation}
so the mixed sector has one-loop topology. Equation~\eqref{eq:intro_complete_order_six}
should therefore be understood as the formal decomposition of the connected
order-six sector, not as an explicit evaluation of its complete momentum and
helicity kernel.

The ultraviolet behavior of $\mathcal{M}_{24}$ can be obtained directly
from the general proposition below. The estimate requires the
Lorentz-invariant phase-space measure, the radiative shear normalization, all
nested principal-value denominators inherited from the lower-order
Yang--Feldman iteration, and the homogeneity of the fourth-order geometrical
kernel. No explicit closed expression for
$\delta a_{4,\lambda}^{\mathrm{out}}$ is needed for this purpose.

%=====================================================================
\section{Ultraviolet power counting beyond third order}
\label{sec:UV_general_order}
%=====================================================================
The ultraviolet counting combines four ingredients: radiative-mode
normalization, canonical contractions, the nested principal-value
denominators in the geometrical vertices, and the independent loop
integrations that remain after imposing the momentum constraints.

The physical quantities considered below are connected scattering
coefficients with $p$ incoming and $q$ outgoing gravitons. The outgoing
operator corrections supply the effective vertices, and canonical
contractions join their incoming-mode factors into internal lines. The
momentum constraints then determine how many independent loop momenta
remain. We first perform this counting for $2\to2$ and then extend it to
arbitrary $p$ and $q\geq2$.

The proposition below applies to the flat-cone sector used throughout the
scattering construction. Its null generators, affine parameter, and
radiative projection are those of the background Minkowski cone. The
explicit inverse-metric correction is excluded from the proposition.

Ultraviolet divergences are a central obstruction in perturbative quantum
gravity. Pure Einstein gravity is finite on shell at one loop but develops a
nonrenormalizable divergence at two
loops~\cite{tHooftVeltman1974,GS86,vanDeVen1992}. This motivates the all-order
uniform radial ultraviolet test of the flat-cone construction carried out
below. The test concerns hard internal momenta at fixed external data.
It does not analyze nonuniform ultraviolet subregions. Infrared and
angular singularities are distinct from the ultraviolet limit considered
here.

In the NSF the physical degrees of freedom are purely radiative: the free
data are the Bondi shear modes on $\scri^\pm$, and the nonradiative sector
is encoded in the nonlinear sources $\Omega_n$ and $\Lambda_n$ rather than in
independent off-shell lines. Every internal line in the counting below is
therefore an on-shell radiative mode, with weight $\dd\omega\,\dd^2\widehat
k/\omega$ (Eq.~\eqref{eq:UV_internal_line_rule}), which is invariant under
a common rescaling of the momenta. The independence of the exponent in
Eq.~\eqref{eq:UV_flat_pq_general_bound} on the loop number follows from this
scale invariance together with the vertex degree $\mathcal O(K)$ assumed in
the proposition. As in Ref.~\cite{KDIV2026}, where the radiative Fock domain is built from
wave packets with smooth profiles supported away from $\omega=0$, only
radiative modes with $\omega>0$ strictly are retained. This restriction
is operative and not only formal, since the same weight $\dd\omega/\omega$ is
logarithmically singular at $\omega\to0$. The estimate below concerns the
ultraviolet end at fixed external data and does not assert finiteness of the
complete amplitude. Likewise, any comparison with perturbative amplitudes
containing soft-graviton divergences refers to the part defined for
$\omega>0$.

%---------------------------------------------------------------------
The shear and oscillator conventions used in this analysis are
\begin{align}
\sigma(u,\Omega)&=\sqrt{4\pi G}\int_0^\infty\frac{d\omega}{2\pi}
\left[a_+(\omega,\Omega)e^{-i\omega u}
+a_-^\dagger(\omega,\Omega)e^{i\omega u}\right],\\
a_\lambda(\omega,\Omega)&:=a_\lambda(\mathbf k=\omega\widehat{\mathbf k}(\Omega)),
\qquad \omega_k=|\mathbf k|,\\
[a_\lambda(\mathbf k),a^\dagger_{\lambda'}(\mathbf k')]
&=\omega_k\delta_{\lambda\lambda'}\delta^{(3)}(\mathbf k-\mathbf k').
\label{eq:main_shear_cartesian_normalization}
\end{align}
The frequency--angle notation denotes the same Cartesian operator, without
rescaling. The conversion to null-cone News modes and the shear commutator
are given in Appendix~\ref{app:shear_null_cone_normalization}.

\subsection{Radiative normalization and canonical contractions}
\label{subsec:UV_normalization_CCR}
%---------------------------------------------------------------------
The asymptotic quantization and canonical commutator were fixed in
Subsection~\ref{subsec:asymptotic_quantization_III}, following
Refs.~\cite{Ashtekar81,Ashtekar1987}. For the ultraviolet counting, the free
Bondi shear carries the frequency--angular measure
\begin{equation}
\dd^2z\,\frac{\dd\omega}{2\pi}
\sqrt{4\pi G}.
\label{eq:UV_original_shear_measure}
\end{equation}
Introducing the Lorentz-invariant on-shell measure
\begin{equation}
\dd\mu(k) = \frac{\dd^3k}{(2\pi)^3 2\omega_k},
\qquad
\omega_k = |\vec k|,
\label{eq:UV_invariant_measure}
\end{equation}
the same radiative-mode measure can be written as
\begin{equation}
\dd^2z\,\frac{\dd\omega}{2\pi}
\sqrt{4\pi G} = \dd\mu(k)\,\nu(\omega),
\label{eq:UV_measure_conversion}
\end{equation}
where
\begin{equation}
\boxed{
\nu(\omega) = 8\pi^2\sqrt{4\pi G}\,\omega^{-1}.
}
\label{eq:UV_nu_definition}
\end{equation}
The factor $\nu(\omega)$ is not an additional measure factor. It is
precisely the weight required to rewrite the original
$\dd\omega\,\dd^2z$ expansion in terms of
$\dd\mu(k)$. In particular, the factor $1/\omega$ contained in the
Lorentz-invariant measure has already been taken into account in deriving
Eq.~\eqref{eq:UV_nu_definition}. The shear is expanded directly in the
operators $a$ and $a^\dagger$, with no additional frequency factor
(Eq.~\eqref{eq:asymptotic_shear_quantization_III}); the factor $\omega$ of
$\dot\sigma$ arises from the $u$ derivative. With the commutator below, this
normalization gives $[\sigma(u,z),\bar\sigma(u',z')]\propto
\int_0^\infty(\dd\omega/\omega)\sin[\omega(u-u')]\,\delta^2(z,z')$, i.e.\ the
canonical step-function commutator of the shear.

The notation
\begin{equation}
a_\lambda(\omega,z)
\equiv
a_\lambda\left(
\vec k = \omega\,\widehat k(z)
\right)
\label{eq:UV_operator_coordinate_notation}
\end{equation}
does not define a rescaled operator. It denotes the same momentum-space
operator expressed in frequency--angular coordinates. The canonical
commutation relation used throughout the calculation is
\begin{equation}
\boxed{
\left[
a_{\lambda}^{\mathrm{in}}(\vec k),
a_{\lambda'}^{\mathrm{in}\dagger}(\vec k')
\right]
=
\omega_k\,\delta_{\lambda\lambda'}\,\delta^{(3)}(\vec k-\vec k').
}
\label{eq:UV_CCR}
\end{equation}
Consider one contraction joining radiative operators belonging to two
different vertices. Before performing the momentum integration, the mode
normalizations and the canonical commutator give
\begin{align}
&\nu(\omega_k)\nu(\omega_{k'})
\left[
a_{\lambda}^{\mathrm{in}}(\vec k),
a_{\lambda'}^{\mathrm{in}\dagger}(\vec k')
\right]
\nonumber \\
&\hspace{15mm}\sim
\frac{\omega_k}{\omega_k\,\omega_{k'}}\,\delta_{\lambda\lambda'}\,\delta^{(3)}(\vec k-\vec k').
\label{eq:UV_CCR_frequency_factors}
\end{align}
After the delta function imposes $\omega_{k'}=\omega_k$, the two mode
normalizations contribute $\omega_k^{-2}$, whereas the commutator
contributes one power of $\omega_k$ in the numerator. Therefore
\begin{equation}
\boxed{
\frac{\omega_k}{\omega_k^2} \sim \frac{1}{\omega_k}.
}
\label{eq:UV_CCR_net_factor}
\end{equation}
Including the invariant measures, the complete contraction has the form
\begin{align}
&\int
\dd\mu(k)\,\dd\mu(k')\,\nu(\omega_k)\nu(\omega_{k'})
\left[
a_{\lambda}^{\mathrm{in}}(\vec k),
a_{\lambda'}^{\mathrm{in}\dagger}(\vec k')
\right]
F(k,k')
\nonumber \\
&\hspace{15mm}\propto
\int
\dd\mu(k)\,\nu(\omega_k)^2 F(k,k),
\label{eq:UV_single_contraction}
\end{align}
where numerical constants independent of the internal momentum have been
suppressed. Radially,
\begin{align}
\dd\mu(k)\,\nu(\omega_k)^2
&\sim
\frac{\dd^3k}{\omega_k}\frac{1}{\omega_k^2}
\nonumber \\
&=
\frac{\dd^3k}{\omega_k^3}
\nonumber \\
&\sim
\boxed{
\frac{\dd\omega_k\,\dd^2\widehat k}{\omega_k}
}.
\label{eq:UV_internal_line_rule}
\end{align}
Equation~\eqref{eq:UV_internal_line_rule} is the complete contribution of
one contracted internal on-shell radiative mode. It includes the surviving
invariant measure, the two factors $\nu(\omega)$, and the factor of
$\omega$ supplied by the canonical commutator. No additional power of
$1/\omega$ from the Lorentz-invariant measure is to be inserted
afterwards. The normalization factors associated with external legs depend
only on fixed external energies and do not affect the ultraviolet degree.

%---------------------------------------------------------------------
\subsection{Flat-cone approximation}
\label{subsec:UV_flat_cone_scope}
%---------------------------------------------------------------------
The scattering amplitudes considered here are extracted using the background
Minkowski null cones and their unperturbed null geodesics. Consistently with
this approximation, the metric contraction in the cone source is evaluated
with the flat inverse metric,
\begin{equation}
g^{ab}\partial_a\Lambda\partial_b\bar{\Lambda}
\longrightarrow
\eta^{ab}\partial_a\Lambda\partial_b\bar{\Lambda}.
\label{eq:UV_flat_metric_replacement}
\end{equation}
The source entering the main proposition is therefore
\begin{equation}
\boxed{
\mathcal{J}_n^{\mathrm{flat}}
=
\left[
\int_s^\infty
\left(
\bar{\eth}(\eth\Omega^2)
+
\eta^{ab}\partial_a\Lambda\partial_b\bar{\Lambda}
\right)
\dd s'
\right]_n.
}
\label{eq:UV_flat_cone_source}
\end{equation}
This sector retains the conformal-factor terms, the mixed
$\Lambda_i\bar{\Lambda}_j$ terms, and all lower-order cuts entering
recursively in those quantities. The term excluded from the main
proposition is the explicit metric insertion
\begin{equation}
\Delta\mathcal{J}_n^{\mathrm{metric}}
=
\left[
\int_s^\infty
\delta g^{ab}\partial_a\Lambda\partial_b\bar{\Lambda}\,\dd s'
\right]_n.
\label{eq:UV_metric_insertion}
\end{equation}
Retaining $\delta g^{ab}$ while keeping the null generators in their
Minkowski form would include only one part of the perturbation of the cone
geometry. A complete metric-corrected construction would also require the
perturbations of the null geodesics, the affine parameter, and the cone
integration measure. For this reason, the explicit metric insertion is not
part of the flat-cone proposition below.

%---------------------------------------------------------------------
\subsection{Nested principal-value denominators and radial degree of the flat-cone vertices}
\label{subsec:UV_flat_vertex_degree}
%---------------------------------------------------------------------
Beginning with $Z_2$, every new retarded--advanced cone integration
produces an outer principal-value factor of the form
\begin{equation}
2i\,\mathrm{PV}\frac{1}{D},
\qquad
D = l(z')\cdot P,
\label{eq:UV_elementary_PV_factor}
\end{equation}
where $P$ is the signed momentum combination associated with the
corresponding branch.

At higher perturbative orders, the lower-order cuts entering recursively in
the source already contain the principal-value denominators generated at the
preceding orders. The new order-$n$ cone integration contributes one
additional factor $2i\,\mathrm{PV}(1/D)$. Consequently, after all lower-order
cuts have been substituted, every complete flat-cone branch of $Z_n$
contains $n-1$ principal-value factors:
\begin{equation}
\boxed{
Z_{n,\mathrm{cone}}^{\mathrm{flat}}
\sim
\mathcal{N}_n^{\mathrm{flat}}
\prod_{\alpha=1}^{n-1}
\left[
2i\,\mathrm{PV}\frac{1}{D_{n,\alpha}}
\right].
}
\label{eq:UV_nested_PV_general}
\end{equation}
Here $\mathcal{N}_n^{\mathrm{flat}}$ denotes the complete momentum
numerator together with angular Green functions, helicity kernels, and
numerical coefficients.

The denominators $D_{n,\alpha}$ generally contain different signed
combinations of the momenta. We therefore define the complete recursive
denominator
\begin{equation}
\boxed{
\mathcal{D}_n
\equiv
\prod_{\alpha=1}^{n-1}D_{n,\alpha}.
}
\label{eq:UV_complete_recursive_denominator}
\end{equation}
Each elementary denominator is linear in its corresponding momentum
combination, but the complete denominator contained in $Z_n$ has radial
degree
\begin{equation}
\boxed{
\mathcal{D}_n
=
\mathcal{O}\left(K^{n-1}\right)
}
\label{eq:UV_complete_denominator_degree}
\end{equation}
under a uniform hard scaling. Equivalently,
\begin{equation}
\prod_{\alpha=1}^{n-1}
\mathrm{PV}\frac{1}{D_{n,\alpha}}
=
\mathcal{O}\left(K^{-(n-1)}\right).
\label{eq:UV_complete_PV_degree}
\end{equation}

The first perturbative orders display the recursive structure explicitly.
At second order,
\begin{equation}
Z_{2,\mathrm{cone}}^{\mathrm{flat}}
\sim
\mathcal{N}_2^{\mathrm{flat}}
\left[
2i\,\mathrm{PV}\frac{1}{D_{2,1}}
\right],
\qquad
\mathcal{D}_2
=
D_{2,1}
=
\mathcal{O}(K).
\label{eq:UV_Z2_denominator_structure}
\end{equation}
At third order,
\begin{equation}
Z_{3,\mathrm{cone}}^{\mathrm{flat}}
\sim
\mathcal{N}_3^{\mathrm{flat}}
\left[
2i\,\mathrm{PV}\frac{1}{D_{3,1}}
\right]
\left[
2i\,\mathrm{PV}\frac{1}{D_{3,2}}
\right],
\label{eq:UV_Z3_denominator_structure}
\end{equation}
with
\begin{equation}
\mathcal{D}_3
=
D_{3,1}D_{3,2}
=
\mathcal{O}(K^2).
\label{eq:UV_D3_degree}
\end{equation}
One denominator is inherited from the internal $Z_2$, while the second is
generated by the new third-order cone integration.

At fourth order,
\begin{equation}
Z_{4,\mathrm{cone}}^{\mathrm{flat}}
\sim
\mathcal{N}_4^{\mathrm{flat}}
\prod_{\alpha=1}^{3}
\left[
2i\,\mathrm{PV}\frac{1}{D_{4,\alpha}}
\right],
\label{eq:UV_Z4_denominator_structure}
\end{equation}
and
\begin{equation}
\mathcal{D}_4
=
D_{4,1}D_{4,2}D_{4,3}
=
\mathcal{O}(K^3).
\label{eq:UV_D4_degree}
\end{equation}
Each subsequent perturbative order adds one further principal-value
denominator.

Consider now the uniform hard scaling
\begin{equation}
k_j^a
\longrightarrow
K k_j^a,
\qquad
K\longrightarrow\infty,
\label{eq:UV_uniform_scaling}
\end{equation}
with fixed nonzero momentum ratios and generic celestial directions.
For each recursive denominator, the leading coefficient in
$D_{n,\alpha}=K d_{n,\alpha}+\Order(1)$ is assumed nonzero. The estimate
is understood on regions where these coefficients remain bounded away
from zero. The momentum
numerator of an order-$n$ flat-cone term has degree
\begin{equation}
\boxed{
\mathcal{N}_n^{\mathrm{flat}}
=
\mathcal{O}(K^n).
}
\label{eq:UV_flat_numerator_degree}
\end{equation}
Combining the numerator with the complete recursive denominator gives
\begin{align}
Z_{n,\mathrm{cone}}^{\mathrm{flat}}
&\sim
\frac{\mathcal{N}_n^{\mathrm{flat}}}{\mathcal{D}_n}
\nonumber \\
&\sim
\frac{K^n}{K^{n-1}}
\nonumber \\
&=
\boxed{
\mathcal{O}(K)
}.
\label{eq:UV_flat_Zn_degree}
\end{align}
We denote the corresponding reduced geometrical vertex by
$\mathcal{V}_n^{\mathrm{flat}}$. Its uniform radial degree is therefore
\begin{equation}
\boxed{
\mathcal{V}_n^{\mathrm{flat}}
=
\mathcal{O}(K).
}
\label{eq:UV_flat_vertex_general}
\end{equation}
The factors $2i$ do not affect the ultraviolet degree. Angular Green
functions and helicity projectors have radial degree zero, while the momentum
powers generated by spacetime and eth derivatives are already included in
$\mathcal{N}_n^{\mathrm{flat}}$. The radiative extraction does not change
the result because the factor generated by $l^a\partial_a Z_n$ is accompanied by
the corresponding on-shell frequency normalization and is homogeneous of
degree zero.

The counting applies when all signed momentum combinations entering
$\mathcal{D}_n$ scale with the common hard parameter $K$. Counting every
independent loop does not, by itself, establish the scaling when only a
subset becomes hard, when hard scales are hierarchically separated, or
when signed hard momenta cancel so that a recursive denominator no longer
grows linearly. Those ultraviolet subregions require separate estimates.
Singular angular configurations and the definition of products of
principal-value kernels are also outside the radial estimate.

%---------------------------------------------------------------------
\subsection{Internal lines and independent loop integrations for
\texorpdfstring{$p\to q$}{p to q} processes}
\label{subsec:UV_loop_counting}
%---------------------------------------------------------------------
Consider the connected two-vertex contribution
\begin{equation}
\mathcal{M}^{(mn)}
=
\left\langle0\right|
\delta a_m^{\mathrm{out}}
\delta a_n^{\mathrm{out}}
a^{\mathrm{in}\dagger}
a^{\mathrm{in}\dagger}
\left|0\right\rangle_{\mathrm{conn}}.
\label{eq:UV_Mmn_definition}
\end{equation}
The operators $\delta a_m^{\mathrm{out}}$ and
$\delta a_n^{\mathrm{out}}$ contain $m$ and $n$ incoming radiative
operators, respectively. Two operator ends are contracted with the two
external incoming creation operators. The remaining $m+n-2$ ends form
internal pairs. Therefore the number of internal on-shell lines is
\begin{equation}
\boxed{
I_{mn}
=
\frac{m+n-2}{2}.
}
\label{eq:UV_Mmn_internal_lines}
\end{equation}
Operator-number parity requires $m+n$ to be even.

For a connected graph containing two effective vertices,
\begin{equation}
\boxed{
L_{mn}
=
I_{mn}-2+1
=
I_{mn}-1
=
\frac{m+n-4}{2}.
}
\label{eq:UV_Mmn_loops}
\end{equation}
After the canonical contractions and vertex constraints are imposed,
$L_{mn}$ independent three-momenta remain.

After the overall external momentum-conservation delta has been factored
out, the internal contribution has the schematic form
\begin{equation}
\int
\left[
\prod_{j=1}^{I_{mn}}
\dd\mu(k_j)\,\nu(\omega_j)^2
\right]
\delta^{(3)}
\left(
\sum_{j=1}^{I_{mn}}\eta_j\vec k_j-\vec P
\right)
\mathcal{V}_m\mathcal{V}_n,
\label{eq:UV_general_internal_integral}
\end{equation}
where $\vec P$ is fixed by the external momenta and
$\eta_j=\pm1$ specifies the momentum branch.

Using the complete internal-line factor
\begin{equation}
\dd\mu(k_j)\,\nu(\omega_j)^2
\sim
\frac{\dd^3k_j}{\omega_j^3},
\label{eq:UV_line_factor_cartesian}
\end{equation}
the uniform radial scaling of the contracted internal phase space is
\begin{align}
&\left[
\prod_{j=1}^{I_{mn}}
\frac{\dd^3k_j}{\omega_j^3}
\right]
\delta^{(3)}
\left(
\sum_{j=1}^{I_{mn}}\eta_j\vec k_j-\vec P
\right)
\nonumber \\
&\hspace{15mm}\sim
K^{3I_{mn}-4} K^{-3I_{mn}}\,\dd K
\nonumber \\
&\hspace{15mm}=
\boxed{
\frac{\dd K}{K^{4}}
}.
\label{eq:UV_contracted_phase_space}
\end{align}
The factor $K^{3I_{mn}-4}\,\dd K$ is the radial volume element of the
$3I_{mn}$-dimensional internal momentum space after the remaining
three-dimensional momentum constraint has been imposed. Equivalently, the
on-shell measures and the momentum constraint contribute
$K^{2I_{mn}-4}\,\dd K$, while the radiative mode normalizations contribute
$K^{-2I_{mn}}$.

Equation~\eqref{eq:UV_contracted_phase_space} contains all independent loop
measures, all internal mode normalizations, and every canonical commutator.
The geometrical vertex factors must be multiplied only after this complete
contracted phase-space contribution has been obtained.

The same counting extends directly to a connected process with $p$
incoming and $q$ outgoing gravitons.  Write the corresponding contribution
as
\begin{equation}
\mathcal{M}_{\{n_\alpha\}}^{(p\to q)}
=
\left\langle0\right|
\prod_{\alpha=1}^{q}
\delta a_{n_\alpha}^{\mathrm{out}}(f_\alpha)
\prod_{\beta=1}^{p}
a^{\mathrm{in}\dagger}(i_\beta)
\left|0\right\rangle_{\mathrm{conn}},
\label{eq:UV_Mpq_definition}
\end{equation}
where $f_\alpha$ and $i_\beta$ collectively denote the momenta and
helicities of the outgoing and incoming external legs.  Since
\(\delta a_{n_\alpha}^{\mathrm{out}}\) contains $n_\alpha$ incoming
radiative operators, the number of internal on-shell lines is
\begin{equation}
\boxed{
I_{\{n_\alpha\}}^{(p,q)}
=
\frac{1}{2}
\left(
\sum_{\alpha=1}^{q}n_\alpha-p
\right).
}
\label{eq:UV_Mpq_internal_lines}
\end{equation}
Thus a nonvanishing Wick sector requires
\(\sum_\alpha n_\alpha-p\) to be a nonnegative even integer.  For a
connected graph with $q$ effective vertices, the number of independent
loop momenta is
\begin{equation}
\boxed{
L_{\{n_\alpha\}}^{(p,q)}
=I_{\{n_\alpha\}}^{(p,q)}-q+1
=
\frac{1}{2}
\left(
\sum_{\alpha=1}^{q}n_\alpha-p-2q+2
\right).
}
\label{eq:UV_Mpq_loops}
\end{equation}

After the overall external momentum-conservation delta has been factored
out, $q-1$ independent three-dimensional vertex constraints remain.
Using Eq.~\eqref{eq:UV_line_factor_cartesian}, the complete internal phase
space therefore scales as
\begin{align}
&\left[
\prod_{j=1}^{I_{\{n_\alpha\}}^{(p,q)}}
\frac{\dd^3k_j}{\omega_j^3}
\right]
\prod_{\rho=1}^{q-1}
\delta^{(3)}
\left(
\sum_j\eta_{\rho j}\vec k_j-\vec P_\rho
\right)
\nonumber \\
&\hspace{15mm}\sim
K^{3I_{\{n_\alpha\}}^{(p,q)}-3q+2}
K^{-3I_{\{n_\alpha\}}^{(p,q)}}\,\dd K
\nonumber \\
&\hspace{15mm}=
\boxed{
\frac{\dd K}
{K^{3q-2}}
}.
\label{eq:UV_Mpq_contracted_phase_space}
\end{align}
Here the momenta \(\vec P_\rho\) are fixed by the external legs and
\(\eta_{\rho j}=0,\pm1\) specifies the momentum routing.  For $p=q=2$,
with $n_1=m$ and $n_2=n$, Eqs.~\eqref{eq:UV_Mpq_internal_lines}--
\eqref{eq:UV_Mpq_contracted_phase_space} reduce to
Eqs.~\eqref{eq:UV_Mmn_internal_lines}--
\eqref{eq:UV_contracted_phase_space}.

%---------------------------------------------------------------------
\subsection{Uniform radial bound in the flat-cone sector}
\label{subsec:UV_flat_proposition}
%---------------------------------------------------------------------
\paragraph{\textbf{Proposition.}}
Assume that the complete order-$n$ flat-cone vertex has the recursive
principal-value structure
\eqref{eq:UV_nested_PV_general}, with
\begin{equation}
\mathcal{N}_n^{\mathrm{flat}}
=
\mathcal{O}(K^n),
\qquad
|\mathcal{D}_n|\geq c\,K^{n-1},\quad c>0.
\label{eq:UV_proposition_hypothesis}
\end{equation}
Under the uniform hard scaling specified above, every connected
$p\to q$ contribution of the form \eqref{eq:UV_Mpq_definition}, with
arbitrary incoming multiplicity $p$, outgoing multiplicity $q\geq2$, and
at least one independent loop momentum, is radially convergent in that
region. This statement holds at arbitrary loop number; it is not a
convergence criterion for all ultraviolet subregions.

Each of the $q$ reduced vertices has the same
uniform degree,
\begin{equation}
\mathcal{V}_{n_\alpha}^{\mathrm{flat}}
=
\mathcal{O}(K),
\qquad
\alpha=1,\ldots,q,
\label{eq:UV_Mpq_flat_vertex_degree}
\end{equation}
and hence
\begin{equation}
\prod_{\alpha=1}^{q}
\mathcal{V}_{n_\alpha}^{\mathrm{flat}}
=
\mathcal{O}(K^q).
\label{eq:UV_Mpq_flat_vertices}
\end{equation}
To display the loop-number dependence, use $I=L+q-1$ and write the
uniform radial count before combining powers. The $L$ independent
three-momenta have radial measure $K^{3L-1}dK$; the $I$ contracted
internal lines contribute $K^{-3I}$, and the $q$ reduced vertices
contribute $K^q$. Thus, on the stated generic hard-momentum region,
\begin{align}
\mathcal M_{\{n_\alpha\},\mathrm{flat}}^{(p\to q),\mathrm{UV}}
&\sim\int^\infty \underbrace{K^{3L-1}\,dK}_{L\text{ loop momenta}}
\underbrace{K^{-3I}}_{I\text{ internal lines}}
\underbrace{K^q}_{q\text{ vertices}}\nonumber\\
&=\int^\infty dK\,K^{3L-1-3(L+q-1)+q}
=\int^\infty\frac{dK}{K^{2q-2}}.
\label{eq:UV_loop_cancellation_explicit}
\end{align}
The $3L$ powers from the independent loop measures cancel the $-3L$
powers from the internal-line weights. This cancellation makes the
radial exponent independent of $L$; it relies on the corrected line
weight $d^3k/\omega^3$ and the assumed vertex degree.

Equivalently, combining this result with
Eq.~\eqref{eq:UV_Mpq_contracted_phase_space} yields
\begin{align}
\mathcal{M}_{\{n_\alpha\},\mathrm{flat}}^{(p\to q),\mathrm{UV}}
&\sim
\int^\infty
\frac{\dd K}
{K^{3q-2}}\,K^q
\nonumber \\
&=
\boxed{
\int^\infty
\frac{\dd K}
{K^{2q-2}}
}.
\label{eq:UV_flat_pq_general_bound}
\end{align}
Since $q\geq2$, the exponent $2q-2$ is at least two, so the radial
integral converges. The exponent does not depend on the loop number, and
the margin of convergence is one power of $K$: any omitted factor growing
like $K$ or faster would spoil it. The uniform radial region therefore
produces no ultraviolet divergence at any loop number within the flat-cone
sector. This conclusion alone does not
exclude counterterms associated with ultraviolet subregions not examined
here.
For $q=1$, the normally ordered contribution contains a single effective
vertex and no loop of the type considered here, so the loop statement is
vacuous.

For the original two-vertex $2\to2$ sector, one sets
$p=q=2$, $n_1=m$, and $n_2=n$.  Then
\begin{equation}
L_{mn}
=
L_{\{m,n\}}^{(2,2)}
=
\frac{m+n-4}{2},
\label{eq:UV_Mmn_loop_specialization}
\end{equation}
and the general bound reduces to
\begin{equation}
\boxed{
\mathcal{M}_{\mathrm{flat}}^{(mn),\mathrm{UV}}
\sim
\int^\infty
\frac{\dd K}{K^{2}}.
}
\label{eq:UV_flat_general_bound}
\end{equation}

\paragraph{\textbf{Finite-order cumulative cut and unitary implementation.}}
The radial proposition and the operator realization address different
questions. The former controls the specified hard-momentum region; it
does not establish self-adjointness or the existence of all products of
principal-value kernels. For the operator construction, we work on a
smeared radiative domain admitting self-adjoint realizations of the
finite-order cuts. With this domain assumption, the truncated operator
\begin{equation}
    \widehat Z_{[N]}
    =
    \sum_{j=1}^{N}\ep^j\widehat Z_j
    \label{eq:UV_finite_truncated_cut_III}
\end{equation}
is taken with a self-adjoint realization and generates the unitary and
bounded operator
\begin{equation}
    \boxed{
    \widehat U_{\omega,[N]}^{\dagger}
    \widehat U_{\omega,[N]}
    =\mathbf 1,
    \qquad
    \|\widehat U_{\omega,[N]}\|=1.
    }
    \label{eq:UV_bounded_cumulative_unitary_III}
\end{equation}
The recursive factors defined in
Eq.~\eqref{eq:incremental_unitary_factor_III} are likewise unitary and have
unit norm. The powers of $\omega$ in the coefficients
$\widehat{\mathcal U}_{\omega,r}$ arise from the expansion of this
operator. Its boundedness does not imply convergence of individual
perturbative coefficients or loop integrals; the stated ultraviolet result
follows from the separate radial power counting.

\paragraph{\textbf{Corollary: the $\mathcal{M}_{33}$ and
$\mathcal{M}_{24}$ sectors.}}
For the third-order diagonal sector,
\begin{equation}
I_{33}=\frac{3+3-2}{2}=2,
\qquad
L_{33}=I_{33}-1=1.
\label{eq:UV_M33_topology_corollary}
\end{equation}
The proposition therefore gives
\begin{equation}
\boxed{
\mathcal{M}_{33}^{\mathrm{flat},\mathrm{UV}}
\sim
\int^\infty\frac{\dd K}{K^2}.
}
\label{eq:UV_M33_corollary}
\end{equation}
For the mixed sector,
\begin{equation}
I_{24}=\frac{2+4-2}{2}=2,
\qquad
L_{24}=I_{24}-1=1,
\label{eq:UV_M24_topology_corollary}
\end{equation}
and hence
\begin{equation}
\boxed{
\mathcal{M}_{24}^{\mathrm{flat},\mathrm{UV}}
\sim
\int^\infty\frac{\dd K}{K^2}.
}
\label{eq:UV_M24_corollary}
\end{equation}
The sectors $\mathcal{M}_{33}$ and $\mathcal{M}_{24}$ therefore have the
same one-loop topology and the same uniform high-energy radial degree. In
particular,
\begin{equation}
\boxed{
\mathcal{M}_{33}^{\mathrm{flat},\mathrm{UV}}
+
\mathcal{M}_{24}^{\mathrm{flat},\mathrm{UV}}
=
\mathcal{O}\!\left(
\int^\infty\frac{\dd K}{K^2}
\right).
}
\label{eq:UV_order_six_sum_corollary}
\end{equation}
Only the operator content of $\delta a_4^{\mathrm{out}}$, the loop count,
and the radial homogeneity of $Z_{4,\mathrm{cone}}^{\mathrm{flat}}$ enter
this conclusion; the explicit fourth-order momentum kernel is not needed.

\section{Discussion and conclusions}
\label{sec:discussion_and_conclusions}
The fourth-order NSF hierarchy identifies the mixed contribution
$\mathcal M_{24}+\mathcal M_{42}$ required alongside $\mathcal M_{33}$
in the two-vertex one-loop four-point sector at absolute perturbative
order six. The outgoing correction is determined formally by the matched
fourth-order source and lower-order cut terms. We have not evaluated its
complete momentum and helicity kernel. Its operator content nevertheless
fixes the mixed one-loop topology, and the recursive source structure
supplies the homogeneity needed for the ultraviolet estimate.

The general fourth-order matching retains the operator-valued phase
$U_\omega$ and the recursive terms it generates. For the scattering
comparison we subsequently set $U_\omega=1$ and apply the simultaneous
Minkowskian approximations of Part II. The extraction geometry remains
classical, while the background phase $e^{-i\omega Z_0}$ and the
nonlinear interaction sources are retained. This separates the general
fourth-order construction from the comparison sector in which the loop
and ultraviolet results are obtained. The technical cut-phase expansion
is given in Appendix~\ref{sec:nsf_yang_feldman}.

The power counting combines the mode normalization and canonical
commutator before assigning a degree to each internal line. Each
contracted on-shell radiative line contributes $d^3k/\omega_k^3$.
The complete recursive flat-cone vertex has uniform hard degree
$\mathcal O(K)$ under the hypotheses of the proposition. For a connected
$p\to q$ sector, the remaining vertex constraints and independent loop
measures then give
\begin{equation}
\mathcal M_{\{n_\alpha\},\mathrm{flat}}^{(p\to q),\mathrm{UV}}
\sim\int^\infty\frac{dK}{K^{2q-2}},
\qquad
L=\frac12\left(\sum_{\alpha=1}^q n_\alpha-p\right)-q+1.
\end{equation}
For arbitrary $p$, $q\geq2$, and $L\geq1$, this integral converges in
the specified uniform radial region. The exponent does not depend on $L$,
and the margin is one power of $K$: omitted factors growing like $K$ or
faster, such as the terms containing $\omega\widehat Z_j$ dropped by the
comparison prescription, would spoil the convergence. In particular, the
mixed sector has the same radial ultraviolet behavior $\int^\infty dK/K^2$
as $\mathcal M_{33}$.

The independence of the uniform radial exponent from $L$ is a feature
of this count. Each additional independent loop momentum contributes a
factor $K^3$, while the corresponding contracted radiative line
contributes $K^{-3}$. This may be compared with the momentum powers
associated with additional loops in conventional perturbative gravity.
The present observation concerns the specified flat-cone sectors and
uniform hard-momentum region.

The result concerns the fixed-background comparison sector of Parts I
and II. Minkowski metric contractions, null trajectories, and background
cuts provide the common setting for comparison with conventional
perturbative graviton scattering. The nonlinear sources remain present;
the approximation does not remove the interaction-generated cut
coefficients. 

The convergence estimate is uniform radial power counting at fixed
external data. It does not establish convergence in ultraviolet regions
with only a subset of hard loop momenta, hierarchically separated hard
scales, or exceptional cancellations in recursive denominators.
Infrared and angular singularities are distinct questions. Boundedness
of a unitary realization of the cut phase also does not replace the
analysis of individual perturbative integrals.

An extension beyond the fixed-Minkowski-background sector requires a
classical fiducial metric in the spacetime interior, where the coordinates
$x^a$ remain classical. The associated null geodesics, cuts, affine
parameter, and cone measure must be constructed consistently with that
metric. We leave this extension to future work; the present calculation
uses the fixed-background sector for comparison with perturbative quantum
gravity.

The next ultraviolet step is to examine the nonuniform hard-momentum
regions of the recursive kernels. Evaluating the complete fourth-order
kernel would additionally determine the momentum and helicity dependence
of the mixed contribution, beyond the topology and radial degree
established here.

\appendix
\section{Detailed matching and recursive cut construction}
\label{sec:nsf_yang_feldman}
%=====================================================================
The prescription used below is the NSF analogue of the causal
Yang--Feldman construction~\cite{YangFeldman}. The local NSF equations are
first solved with retarded boundary conditions and therefore determine all
interior fields from the incoming radiative data. The same local solution is
then inserted in the future and antipodally identified past cone
representations. The two cone terms differ only in the orientation of the
outer affine integration. A nonlinear scalar model illustrating this logic is
given in Appendix~\ref{app:scalar_yang_feldman}.

\subsection{Exact equations and perturbative fields}
\label{subsec:exact_nsf_and_expansion}
We use the conventions and normalizations of the first two papers. The total
cut is decomposed into the Minkowski cut and its deformation,
\begin{equation}
Z(x,z)
\equiv
Z_{\mathrm{total}}(x,z)
=
Z_0(x,z)+\mathcal Z(x,z),
\qquad
\mathcal Z
=
\sum_{n\geq1}\ep^n Z_n,
\label{eq:Z_sigma_expansion_III}
\end{equation}
where $Z_0(x,z)=x^a l_a(z)$. The remaining NSF fields are expanded as
\begin{align}
\sigma(u,z)
&=
\sum_{n\geq1}\ep^n\sigma_n(u,z),
&
\bar\sigma(u,z)
&=
\sum_{n\geq1}\ep^n\bar\sigma_n(u,z),
\\
\Lambda
&=
\eth^2 Z
=
\sum_{n\geq1}\ep^n\Lambda_n,
&
\Omega
&=
1+\sum_{n\geq2}\ep^n\Omega_n.
\label{eq:Lambda_Omega_expansion_III}
\end{align}
The dependence of the asymptotic shear on the cut must be kept explicit. The
exact cut equation is
\begin{equation}
\boxed{
\begin{aligned}
\ethb^2\eth^2 Z(x,z)
={}&
\eth^2\bar\sigma\left(Z(x,z),z\right)
+
\ethb^2\sigma\left(Z(x,z),z\right)
-
\Sigma^+(x,z)
\\
&-
\int_r^\infty
\left[
\Omega^{-2}\ethb(\eth\Omega^2)
+
h^{ab}\p_a\Lambda\p_b\bar\Lambda
\right]\dd r'.
\end{aligned}
}
\label{eq:exact_Z_equation_III}
\end{equation}
The flux term is likewise evaluated on the cut,
\begin{equation}
\Sigma^+(x,z)
=
\int_{-\infty}^{Z(x,z)}
\dot\sigma(u,z)\dot{\bar\sigma}(u,z)\,\dd u.
\label{eq:Sigma_definition_III}
\end{equation}
The occurrence of the full cut in the shear argument is essential. Note that the shear terms appearing in
Eq.~\eqref{eq:exact_Z_equation_III} are also evaluated on the cut:
\begin{equation}
\sigma(u,z)\big|_{u=Z(x,z)}
=
\sigma\left(Z(x,z),z\right),
\qquad
\bar\sigma(u,z)\big|_{u=Z(x,z)}
=
\bar\sigma\left(Z(x,z),z\right).
\label{eq:shear_evaluated_on_cut_III}
\end{equation}
This dependence becomes particularly transparent by decomposing the classical
radiative data into positive- and negative-frequency modes. We write
\begin{align}
\sigma(u,z)
={}&
\int_0^\infty
\frac{\dd\omega}{2\pi}
\sqrt{4\pi G}
\left[
\alpha_+(\omega,z)\ee^{-\ii\omega u}
+
\alpha_-^*(\omega,z)\ee^{+\ii\omega u}
\right],
\\
\bar\sigma(u,z)
={}&
\int_0^\infty
\frac{\dd\omega}{2\pi}
\sqrt{4\pi G}
\left[
\alpha_-(\omega,z)\ee^{-\ii\omega u}
+
\alpha_+^*(\omega,z)\ee^{+\ii\omega u}
\right].
\label{eq:classical_shear_Fourier_III}
\end{align}
Here $\alpha_+(\omega,z)$ and $\alpha_-(\omega,z)$ are classical
positive-frequency radiative amplitudes of helicity $+2$ and $-2$,
respectively. Their complex conjugates multiply the corresponding
negative-frequency phases. No quantization has yet been performed.
Evaluating the radiative data on the cut,
\begin{equation}
u=Z(x,z),
\end{equation}
gives
\begin{align}
\sigma\left(Z(x,z),z\right)
={}&
\int_0^\infty
\frac{\dd\omega}{2\pi}
\sqrt{4\pi G}
\left[
\alpha_+(\omega,z)\ee^{-\ii\omega Z(x,z)}
+
\alpha_-^*(\omega,z)\ee^{+\ii\omega Z(x,z)}
\right],
\\
\bar\sigma\left(Z(x,z),z\right)
={}&
\int_0^\infty
\frac{\dd\omega}{2\pi}
\sqrt{4\pi G}
\left[
\alpha_-(\omega,z)\ee^{-\ii\omega Z(x,z)}
+
\alpha_+^*(\omega,z)\ee^{+\ii\omega Z(x,z)}
\right].
\label{eq:classical_shear_on_cut_Fourier_III}
\end{align}
Thus, the dependence on the complete nonlinear cut appears directly through
the two phases
\begin{equation}
\ee^{-\ii\omega Z(x,z)},
\qquad
\ee^{+\ii\omega Z(x,z)}.
\label{eq:full_cut_two_phases_III}
\end{equation}
Separating the Minkowski cut from its nonlinear deformation,
\begin{equation}
Z(x,z)=Z_0(x,z)+\mathcal Z(x,z),
\label{eq:flat_nonlinear_cut_split_III}
\end{equation}
the positive-frequency phase becomes
\begin{equation}
\ee^{-\ii\omega Z(x,z)}
=
\ee^{-\ii\omega Z_0(x,z)}
\ee^{-\ii\omega\mathcal Z(x,z)},
\label{eq:positive_frequency_cut_factorization_III}
\end{equation}
while the negative-frequency phase is
\begin{equation}
\ee^{+\ii\omega Z(x,z)}
=
\ee^{+\ii\omega Z_0(x,z)}
\ee^{+\ii\omega\mathcal Z(x,z)}.
\label{eq:negative_frequency_cut_factorization_III}
\end{equation}
We therefore define the classical cut factors
\begin{equation}
\boxed{
U_\omega[\mathcal Z]
\equiv
\exp\left[-\ii\omega\mathcal Z(x,z)\right],
\qquad
U_\omega^*[\mathcal Z]
\equiv
\exp\left[+\ii\omega\mathcal Z(x,z)\right].
}
\label{eq:classical_cut_factors_III}
\end{equation}
The shear evaluated on the cut can then be written as
\begin{align}
\sigma(Z,z)
={}&
\int_0^\infty
\frac{\dd\omega}{2\pi}
\sqrt{4\pi G}
\left[
\alpha_+(\omega,z)\ee^{-\ii\omega Z_0} U_\omega
+
\alpha_-^*(\omega,z)\ee^{+\ii\omega Z_0} U_\omega^*
\right],
\\
\bar\sigma(Z,z)
={}&
\int_0^\infty
\frac{\dd\omega}{2\pi}
\sqrt{4\pi G}
\left[
\alpha_-(\omega,z)\ee^{-\ii\omega Z_0} U_\omega
+
\alpha_+^*(\omega,z)\ee^{+\ii\omega Z_0} U_\omega^*
\right].
\label{eq:classical_shear_on_cut_with_U_III}
\end{align}
Hence the helicity amplitudes, the frequency integral, and the classical
factors $U_\omega$ all arise directly from the Fourier decomposition of the
radiative shear before quantization.
Thus the phase containing the complete nonlinear cut is generated directly
by the Fourier representation of the asymptotic shear.

The factors $U_\omega[\mathcal Z]$ therefore arise before quantization,
solely from evaluating the Fourier modes of the classical shear on the
nonlinear cut. Their perturbative expansion,
\begin{equation}
U_\omega[\mathcal Z]
=
1-\ii\omega\mathcal Z
-\frac{\omega^2}{2}\mathcal Z^2+\cdots,
\label{eq:classical_U_perturbative_expansion_III}
\end{equation}
subsequently generates the familiar Taylor terms involving lower-order
cuts. The Fourier representation will be
taken as the starting point of the matching construction.
Using $\dd r=\Omega^2\dd s$, the radial source is written in terms of the
affine parameter $s$ as
\begin{equation}
\int_s^\infty
\left[
\ethb(\eth\Omega^2)
+
g^{ab}\p_a\Lambda\p_b\bar\Lambda
\right]\dd s'.
\label{eq:affine_cone_source_III}
\end{equation}
We define the order-$n$ generator source by
\begin{equation}
\boxed{
\mathcal J_n(x,z')
\equiv -
\left[
\int_s^\infty
\left(
\ethb(\eth\Omega^2)
+
g^{ab}\p_a\Lambda\p_b\bar\Lambda
\right)\dd s'
\right]_n.
}
\label{eq:In_definition_III}
\end{equation}

At perturbative order $(n)$, the angular NSF equation is inverted with the
scalar Green function $G_{0,0'}$. The complete cut is therefore
\begin{equation}
\boxed{
\begin{aligned}
Z_n(x,z)
&=
\oint \dd^2z'
G_{0,0'}(z,z')
\Big[
\bar\eth'^{2}
\sigma_n
\left(
Z_{\mathrm{total}}(x,z'),z'
\right)
+
\eth'^{2}
\bar\sigma_n
\left(
Z_{\mathrm{total}}(x,z'),z'
\right)
\\
&\quad
-
\Sigma_n
\left(
Z_{\mathrm{total}}(x,z'),z'
\right)
+
\mathcal J_n(x,z')
\Big].
\end{aligned}
}
\label{eq:Zn_complete_angular_inversion_III}
\end{equation}
 Equation~\eqref{eq:Zn_complete_angular_inversion_III}
separates naturally as
\begin{equation}
\boxed{
Z_n
=
Z_{n,\mathrm{cut}}
+
Z_{n,\mathrm{cone}},
}
\label{eq:Zn_cut_cone_split_III}
\end{equation}
where
\begin{equation}
\boxed{
\begin{aligned}
Z_{n,\mathrm{cut}}(x,z)
&=
\oint\dd^2z'
G_{0,0'}(z,z')
\Big[
\bar\eth'^{2}
\sigma_n
\left(
Z_{\mathrm{total}}(x,z'),z'
\right)
+
\eth'^{2}
\bar\sigma_n
\left(
Z_{\mathrm{total}}(x,z'),z'
\right)
\\
&\quad
-
\Sigma_n
\left(
Z_{\mathrm{total}}(x,z'),z'
\right)
\Big],
\end{aligned}
}
\label{eq:Zn_cut_before_parts_III}
\end{equation}
and
\begin{equation}
\boxed{
Z_{n,\mathrm{cone}}(x,z)
=
\oint\dd^2z'
G_{0,0'}(z,z')
\mathcal J_n(x,z').
}
\label{eq:Zn_cone_G00_III}
\end{equation}
Moving the angular derivatives from the radiative data to the Green
function and defining
\begin{equation}
G_{0,-2'}(z,z')
\equiv
\bar\eth'^{2}G_{0,0'}(z,z'),
\qquad
G_{0,2'}(z,z')
\equiv
\eth'^{2}G_{0,0'}(z,z'),
\label{eq:G0_spin_kernels_III}
\end{equation}
the cut contribution becomes
\begin{equation}
\boxed{
\begin{aligned}
Z_{n,\mathrm{cut}}(x,z)
&=
\oint\dd^2z'
\Big[
G_{0,-2'}(z,z')
\sigma_n
\left(
Z_{\mathrm{total}}(x,z'),z'
\right)
+
G_{0,2'}(z,z')
\bar\sigma_n
\left(
Z_{\mathrm{total}}(x,z'),z'
\right)
\\
&\quad
-
G_{0,0'}(z,z')
\Sigma_n
\left(
Z_{\mathrm{total}}(x,z'),z'
\right)
\Big].
\end{aligned}
}
\label{eq:Zn_cut_after_parts_III}
\end{equation}
Thus the two parts of the cut are displayed explicitly: the spin-weighted
radiative contribution is inverted with $G_{0,\pm2'}$, whereas the scalar
flux contribution is inverted directly with $G_{0,0'}$.

The first two terms in
Eq.~\eqref{eq:Zn_cut_after_parts_III} contain the new order-(n)
radiative coefficients, together with contributions determined entirely
by lower-order cuts through the perturbative coefficients of
\(U_\omega\). The scalar flux term is likewise fixed by lower-order
radiative data and does not introduce an additional order-(n)
unknown.
 
The conformal factor satisfies
\begin{equation}
2\p_r^2\Omega=R_{rr}[h]\Omega,
\label{eq:Omega_exact_III}
\end{equation}
where
\begin{equation}
R_{rr}[h]
=
\frac{1}{4q}\p_r^2\Lambda\p_r^2\bar\Lambda
+
\frac{3}{8q^2}(\p_rq)^2
-
\frac{1}{4q}\p_r^2q,
\qquad
q=1-\p_r\Lambda\p_r\bar\Lambda.
\label{eq:Rrr_q_III}
\end{equation}
Once the lower-order cuts are known, the radial equation gives
\begin{equation}
\boxed{
\Omega_n(r)
=
\frac12
\int_r^\infty\dd r'
\int_{r'}^\infty\dd r''\,
\left[R_{rr}\Omega\right]_n(r'').
}
\label{eq:Omega_n_general_III}
\end{equation}
After this radial equation has been solved, $\mathcal J_n$ is a functional of
fields already determined at lower orders. The local equations themselves
carry no future or past label; that distinction enters only through the
asymptotic representation and the orientation of the final cone integral.

\subsection{Incoming representation of the matching equation}
\label{subsec:incoming_matching_representation}
We derive the incoming representation of the matching equation first at the
classical level. Future quantities are kept in their natural retarded-time
representation at $\mathscr I^+$. Past quantities, naturally defined at
$\mathscr I^-$ in terms of the advanced Bondi time $v$, are antipodally
mapped to the future celestial sphere and written in the same $u$
representation by setting $v=-u$. Thus, for any past asymptotic field
$F_{\rm nat}^{-}(v,z)$, we define
\begin{equation}
\boxed{
F^{-}(u,z)
\equiv
\left.
\bigl(\Amap F_{\rm nat}^{-}\bigr)(v,z)
\right|_{v=-u}
=
F_{\rm nat}^{-}(-u,\widehat z).
}
\label{eq:past_to_common_u_representation_III}
\end{equation}
For a field of nonzero spin weight, the appropriate antipodal spin phase is
understood as part of the action of $\Amap$. No such phase occurs for the
cut, which is a scalar on the celestial sphere. In particular,
\begin{equation}
\boxed{
Z_{\rm total}^{-}(x,z)
\equiv
\left.
\bigl(\Amap Z_{\rm total,nat}^{-}\bigr)(x,z)
\right|_{v=-u}
}
\label{eq:incoming_total_cut_definition_III}
\end{equation}
denotes the past total cut in the common $u$ representation. With this
convention,
\begin{equation}
\boxed{
Z_{\rm total}^{+}(x,z)+Z_{\rm total}^{-}(x,z)=0,
}
\label{eq:original_total_matching_III}
\end{equation}
and at linear order
\begin{equation}
\boxed{
Z_0^{+}(x,z)+Z_0^{-}(x,z)=0.
}
\label{eq:linear_cut_matching_common_u_III}
\end{equation}
The part of the future cut generated explicitly by the asymptotic shear is
\begin{equation}
\begin{aligned}
Z_{\sigma}^{+}(x,z)
={}&
\oint\dd^2z' G_{0,0'}(z,z')
\Big[
\bar\eth'^2\sigma_{\rm out}^{+}
\left(Z_{\rm total}^{+}(x,z'),z'\right)
\\
&\hspace{34mm}
+\eth'^2\bar\sigma_{\rm out}^{+}
\left(Z_{\rm total}^{+}(x,z'),z'\right)
\Big].
\end{aligned}
\label{eq:future_shear_cut_G00_III}
\end{equation}
After integrating by parts on the sphere, define the spin-weighted kernels
\begin{equation}
\boxed{
G_{0,-2'}(z,z')
\equiv
\bar\eth'^2G_{0,0'}(z,z'),
\qquad
G_{0,2'}(z,z')
\equiv
\eth'^2G_{0,0'}(z,z').
}
\label{eq:spin_Green_kernels_III}
\end{equation}
The angularly integrated cut is therefore
\begin{equation}
\boxed{
\begin{aligned}
Z_{\sigma}^{+}(x,z)
={}&
\oint\dd^2z'
\Big[
G_{0,-2'}(z,z')
\sigma_{\rm out}^{+}
\left(Z_{\rm total}^{+}(x,z'),z'\right)
\\
&\hspace{29mm}
+
G_{0,2'}(z,z')
\bar\sigma_{\rm out}^{+}
\left(Z_{\rm total}^{+}(x,z'),z'\right)
\Big].
\end{aligned}
}
\label{eq:future_shear_cut_integrated_III}
\end{equation}
In Ref.~\cite{BKR2023}, the linear antipodal matching was written explicitly
for one of the two complex helicity sectors, while the conjugate sector was
contained in the complex-conjugate contribution to the real cut. Restoring
both helicities gives
\begin{equation}
\boxed{
\begin{aligned}
\sigma_{1,\mathrm{out}}^{+}(u,z)
&=
-\bar\sigma_{1,\mathrm{in}}^{-}(u,\widehat z),
\\
\bar\sigma_{1,\mathrm{out}}^{+}(u,z)
&=
-\sigma_{1,\mathrm{in}}^{-}(u,\widehat z).
\end{aligned}
}
\label{eq:linear_two_helicity_matching_III}
\end{equation}
The second relation is the complex conjugate of the first, but both are
shown because they identify separately the two helicity sectors.
For $\omega>0$, the outgoing annihilation branches are
\begin{align}
\sigma_{1,\mathrm{out}}^{+}(u,z')\big|_{\rm ann}
&=
\int_0^\infty\frac{\dd\omega}{2\pi}
\sqrt{4\pi G_{\rm N}}
\alpha_{1,+}^{\rm out}(\omega,z')
\ee^{-\ii\omega u},
\\
\bar\sigma_{1,\mathrm{out}}^{+}(u,z')\big|_{\rm ann}
&=
\int_0^\infty\frac{\dd\omega}{2\pi}
\sqrt{4\pi G_{\rm N}}
\alpha_{1,-}^{\rm out}(\omega,z')
\ee^{-\ii\omega u}.
\label{eq:linear_outgoing_annihilation_branches_III}
\end{align}
Consequently,
\begin{equation}
\boxed{
\begin{aligned}
Z_{\sigma,1}^{+}(x,z)
={}&
\oint\dd^2z'\int_0^\infty\frac{\dd\omega}{2\pi}
\sqrt{4\pi G_{\rm N}}
\Big[
\big(
G_{0,-2'}\alpha_{1,+}^{\rm out}
+
G_{0,2'}\alpha_{1,-}^{\rm out}
\big)
\ee^{-\ii\omega Z_0^{+}}
+\mathrm{c.c.}
\Big].
\end{aligned}
}
\label{eq:linear_future_shear_cut_III}
\end{equation}
The Green functions have arguments $(z,z')$, the amplitudes have arguments
$(\omega,z')$, and the cuts are evaluated at $(x,z')$.
Keeping the fixed minus sign in
Eq.~\eqref{eq:linear_two_helicity_matching_III} explicitly, the
corresponding past contribution is
\begin{equation}
\boxed{
\begin{aligned}
Z_{\sigma,1}^{-}(x,z)
={}&
\oint\dd^2z'\int_0^\infty\frac{\dd\omega}{2\pi}
\sqrt{4\pi G_{\rm N}}
\Big[
\big(
G_{0,-2'}\alpha_{1,+}^{\rm in}
+
G_{0,2'}\alpha_{1,-}^{\rm in}
\big)
\ee^{-\ii\omega Z_0^{-}}
+\mathrm{c.c.}
\Big].
\end{aligned}
}
\label{eq:linear_incoming_shear_cut_III}
\end{equation}
Equivalently, this fixed sign may be absorbed in the definition of the
antipodally identified incoming amplitudes.
Since
\begin{equation}
\boxed{
Z_0^{+}(x,z')=-Z_0^{-}(x,z'),
}
\label{eq:linear_Minkowski_cut_matching_III}
\end{equation}
the two Fourier branches satisfy
\begin{equation}
\boxed{
\ee^{-\ii\omega Z_0^{+}}
=
\ee^{+\ii\omega Z_0^{-}},
\qquad
\ee^{+\ii\omega Z_0^{+}}
=
\ee^{-\ii\omega Z_0^{-}}.
}
\label{eq:linear_phase_matching_III}
\end{equation}
The second equality rewrites the complex-conjugate branch of the future
cut in the same incoming phase used for the past contribution. Expressing
both real cuts in this common incoming representation and defining
\begin{equation}
\delta\alpha_{1,\lambda}
\equiv
\alpha_{1,\lambda}^{\rm out}
-
\alpha_{1,\lambda}^{\rm in},
\qquad
\lambda=\pm,
\label{eq:linear_delta_alpha_definition_III}
\end{equation}
the linear matching equation becomes
\begin{equation}
\boxed{
\begin{aligned}
0
={}&
\oint\dd^2z'\int_0^\infty\frac{\dd\omega}{2\pi}
\sqrt{4\pi G_{\rm N}}
\Big[
\big(
G_{0,-2'}(z,z')
\delta\alpha_{1,+}(\omega,z')
+
G_{0,2'}(z,z')
\delta\alpha_{1,-}(\omega,z')
\big)
\\
&\hspace{46mm}\times
\ee^{-\ii\omega Z_0^{-}(x,z')}
+\mathrm{c.c.}
\Big].
\end{aligned}
}
\label{eq:linear_matching_delta_alpha_III}
\end{equation}
The spin weight of each Green function is opposite to that of the
corresponding helicity amplitude. The two kernels therefore project
separately onto the two helicity sectors, and hence
\begin{equation}
\boxed{
\delta\alpha_{1,+}=0,
\qquad
\delta\alpha_{1,-}=0.
}
\label{eq:linear_delta_alpha_zero_III}
\end{equation}
The same argument applies to the complete cuts. Since
\begin{equation}
\boxed{
Z_{\rm total}^{+}(x,z')
=
-Z_{\rm total}^{-}(x,z'),
}
\label{eq:exact_total_cut_matching_III}
\end{equation}
the full phases satisfy
\begin{equation}
\boxed{
\ee^{-\ii\omega Z_{\rm total}^{+}}
=
\ee^{+\ii\omega Z_{\rm total}^{-}},
\qquad
\ee^{+\ii\omega Z_{\rm total}^{+}}
=
\ee^{-\ii\omega Z_{\rm total}^{-}}.
}
\label{eq:exact_total_phase_matching_III}
\end{equation}
Defining
\begin{equation}
\delta\alpha_\lambda
\equiv
\alpha_\lambda^{\rm out}
-
\alpha_\lambda^{\rm in},
\label{eq:exact_classical_delta_alpha_III}
\end{equation}
the exact shear-dependent cut difference, written in the common incoming
branch, is
\begin{equation}
\boxed{
\begin{aligned}
\Delta Z_\sigma(x,z)
={}&
\oint\dd^2z'\int_0^\infty\frac{\dd\omega}{2\pi}
\sqrt{4\pi G_{\rm N}}
\Big[
\big(
G_{0,-2'}\delta\alpha_+
+
G_{0,2'}\delta\alpha_-
\big)
\ee^{-\ii\omega Z_{\rm total}^{-}}
+\mathrm{c.c.}
\Big].
\end{aligned}
}
\label{eq:exact_classical_incoming_matching_III}
\end{equation}
Only now do we split
\begin{equation}
Z_{\rm total}^{-}
=
Z_0^{-}
+
\mathcal Z^{-},
\label{eq:incoming_total_cut_split_III}
\end{equation}
and define
\begin{equation}
\boxed{
U_\omega^{-}
\equiv
\exp[-\ii\omega\mathcal Z^{-}],
\qquad
\bar U_\omega^{-}
\equiv
\exp[+\ii\omega\mathcal Z^{-}].
}
\label{eq:classical_incoming_U_III}
\end{equation}
Thus,
\begin{equation}
\ee^{-\ii\omega Z_{\rm total}^{-}}
=
\ee^{-\ii\omega Z_0^{-}}U_\omega^{-},
\qquad
\ee^{+\ii\omega Z_{\rm total}^{-}}
=
\bar U_\omega^{-}\ee^{+\ii\omega Z_0^{-}},
\label{eq:incoming_phase_factorization_III}
\end{equation}
and
\begin{equation}
\boxed{
\begin{aligned}
\Delta Z_\sigma(x,z)
={}&
\oint\dd^2z'\int_0^\infty\frac{\dd\omega}{2\pi}
\sqrt{4\pi G_{\rm N}}
\Big[
\big(
G_{0,-2'}\delta\alpha_+
+
G_{0,2'}\delta\alpha_-
\big)
\ee^{-\ii\omega Z_0^{-}}U_\omega^{-}
+\mathrm{c.c.}
\Big].
\end{aligned}
}
\label{eq:exact_classical_incoming_matching_with_U_III}
\end{equation}
\subsection{Asymptotic quantization and the complete cut term}
\label{subsec:asymptotic_quantization_III}
We now quantize the incoming radiative phase space following
Ashtekar~\cite{Ashtekar81,Ashtekar1987}. In the common $u$ representation,
the free incoming shears are
\begin{align}
\widehat\sigma_{\rm in}(u,z)
&=
\int_0^\infty\frac{\dd\omega}{2\pi}
\sqrt{4\pi G_{\rm N}}
\left[
a_{+}^{\rm in}(\omega,z)\ee^{-\ii\omega u}
+
a_{-}^{\rm in\dagger}(\omega,z)\ee^{+\ii\omega u}
\right],
\\
\widehat{\bar\sigma}_{\rm in}(u,z)
&=
\int_0^\infty\frac{\dd\omega}{2\pi}
\sqrt{4\pi G_{\rm N}}
\left[
a_{-}^{\rm in}(\omega,z)\ee^{-\ii\omega u}
+
a_{+}^{\rm in\dagger}(\omega,z)\ee^{+\ii\omega u}
\right].
\label{eq:asymptotic_shear_quantization_III}
\end{align}
Here $a_\lambda(\omega,z)$ denotes the momentum-space operator evaluated at
$\vec k=\omega\widehat k(z)$. The canonical commutator is
\begin{equation}
\boxed{
\left[
a_\lambda^{\rm in}(\vec k),
a_{\lambda'}^{\rm in\dagger}(\vec k')
\right]
=
\omega_k
\delta_{\lambda\lambda'}
\delta^{(3)}(\vec k-\vec k'),
}
\label{eq:asymptotic_CCR_III}
\end{equation}
with all other commutators vanishing. The outgoing operators are written as
\begin{equation}
a_\lambda^{\rm out}
=
a_\lambda^{\rm in}
+
\delta a_\lambda^{\rm out}.
\label{eq:out_in_delta_operator_III}
\end{equation}
Promoting the nonlinear incoming cut to a self-adjoint operator, define
\begin{equation}
\boxed{
\widehat U_\omega^{-}
\equiv
\exp[-\ii\omega\widehat{\mathcal Z}^{-}],
\qquad
\widehat U_\omega^{-\dagger}
=
\exp[+\ii\omega\widehat{\mathcal Z}^{-}].
}
\label{eq:quantum_incoming_U_III}
\end{equation}
The shear-dependent part of the matched quantum cut is then
\begin{equation}
\boxed{
\begin{aligned}
\Delta\widehat Z_\sigma(x,z)
={}&
\oint\dd^2z'\int_0^\infty\frac{\dd\omega}{2\pi}
\sqrt{4\pi G_{\rm N}}
\Big[
\big(
G_{0,-2'}\delta a_+^{\rm out}
+
G_{0,2'}\delta a_-^{\rm out}
\big)
\ee^{-\ii\omega Z_0^{-}}
\widehat U_\omega^{-}
+
\mathrm{H.c.}
\Big].
\end{aligned}
}
\label{eq:exact_matched_quantum_cut_III}
\end{equation}
The Hermitian-conjugate term has the reversed operator order.
The scalar flux enters the cut with the minus sign in
Eq.~\eqref{eq:exact_Z_equation_III}. Define
the normal-ordered, self-adjoint flux density
\begin{equation}
\widehat{\mathcal F}^{\pm}(u,z)
\equiv
{}:
\dot{\widehat\sigma}^{\pm}(u,z)
\dot{\widehat{\bar\sigma}}^{\pm}(u,z)
::{},
\qquad
\partial_u\widehat\Sigma^{\pm}
=
\widehat{\mathcal F}^{\pm}.
\label{eq:quantum_flux_density_III}
\end{equation}
With the Fourier convention
\begin{equation}
\widehat{\mathcal F}^{\pm}(u,z)
=
\int_{-\infty}^{+\infty}
\frac{\dd\omega}{2\pi}
\widehat{\mathcal F}_{\omega}^{\pm}(z)
\ee^{-\ii\omega u},
\label{eq:flux_Fourier_transform_III}
\end{equation}
the retarded integral from $-\infty$ gives
\begin{equation}
\boxed{
\widehat\Sigma_{\omega}^{\pm}(z)
=
\frac{
\widehat{\mathcal F}_{\omega}^{\pm}(z)
}{
-\ii(\omega+\ii0)
}.
}
\label{eq:Sigma_Fourier_transform_III}
\end{equation}
After antipodal matching, define
\begin{equation}
\delta\widehat\Sigma_{\omega}(z)
\equiv
\widehat\Sigma_{\omega,\rm out}(z)
-
\widehat\Sigma_{\omega,\rm in}(z).
\label{eq:delta_Sigma_definition_III}
\end{equation}
Its signed contribution to the cut is
\begin{equation}
\boxed{
\begin{aligned}
\Delta\widehat Z_\Sigma(x,z)
={}&-
\oint\dd^2z'
G_{0,0'}(z,z')
\int_0^\infty\frac{\dd\omega}{2\pi}
\Big[
\delta\widehat\Sigma_{\omega}(z')
\ee^{-\ii\omega Z_0^{-}(x,z')}
\widehat U_\omega^{-}(x,z')
+
\mathrm{H.c.}
\Big].
\end{aligned}
}
\label{eq:exact_matched_flux_cut_III}
\end{equation}
Thus the complete cut difference is
\begin{equation}
\boxed{
\Delta\widehat Z_{\rm cut}
=
\Delta\widehat Z_\sigma
+
\Delta\widehat Z_\Sigma.
}
\label{eq:complete_quantum_cut_difference_III}
\end{equation}
The flux term begins quadratically. At absolute order $n$ its density has
the triangular form
\begin{equation}
\widehat{\mathcal F}_n
=
\sum_{p=1}^{n-1}
{}:
\dot{\widehat\sigma}_p
\dot{\widehat{\bar\sigma}}_{n-p}
::{},
\label{eq:flux_triangular_order_III}
\end{equation}
so it contains no shear of order $n$. Hence it never defines the new
operator $\delta a_{n,\lambda}^{\rm out}$ and is completely determined by
lower-order fields. Moreover, the free quadratic pieces coincide by the
linear matching, so
\begin{equation}
\boxed{
\delta\widehat\Sigma_{\omega,2}=0.
}
\label{eq:delta_Sigma2_zero_III}
\end{equation}
The first potentially nonvanishing flux difference is therefore of third
order. Since it is evaluated on the complete incoming cut, its order-$n$
coefficient is obtained directly from
\begin{equation}
\boxed{
\left[
\delta\widehat\Sigma_{\omega}
\widehat U_\omega^{-}
\right]_n
=
\sum_{r=2}^{n}
\delta\widehat\Sigma_{\omega,r}
\widehat U_{\omega,n-r},
}
\label{eq:Sigma_U_order_expansion_III}
\end{equation}
with no expansion of the upper integration limit.

\subsection{Retarded local fields and the common cone source}
\label{subsec:future_past_cone_maps}
The metric entering the cone equation is locally reconstructed from the NSF
fields,
\begin{equation}
h^{ab}=h^{ab}[\Lambda,\bar\Lambda,\Omega].
\label{eq:metric_local_NSF_fields_III}
\end{equation}
Neither the local Einstein equations nor the perturbative equations for
$\Lambda$ and $\Omega$ distinguish the future reconstruction from the
antipodally identified past reconstruction. Our Yang--Feldman working
prescription is to select the retarded perturbative solution determined by
the incoming free field and to use that solution inside both cone sources.
For every $j<n$,
\begin{align}
\Lambda_j
&\longrightarrow
\Lambda_j[Z^-]
=
\Lambda_j[a^{\mathrm{in}},a^{\mathrm{in}\dagger}],
\\
\bar\Lambda_j
&\longrightarrow
\bar\Lambda_j[Z^-]
=
\bar\Lambda_j[a^{\mathrm{in}},a^{\mathrm{in}\dagger}],
\\
\Omega_j
&\longrightarrow
\Omega_j[Z^-]
=
\Omega_j[a^{\mathrm{in}},a^{\mathrm{in}\dagger}],
\\
h_j^{ab}
&\longrightarrow
h_j^{ab}[Z^-]
=
h_j^{ab}[a^{\mathrm{in}},a^{\mathrm{in}\dagger}].
\label{eq:YF_local_field_replacement_III}
\end{align}
The order-$n$ source is consequently a single operator-valued functional,
\begin{equation}
\boxed{
\widehat{\mathcal J}_n^{\mathrm{YF}}
\equiv
\widehat{\mathcal J}_n
[a^{\mathrm{in}},a^{\mathrm{in}\dagger}].
}
\label{eq:common_incoming_source_n}
\end{equation}
At second order this prescription is automatic. At third order it was
implemented by replacing the lower-order fields by their incoming-data
expressions. Here it is imposed explicitly at fourth and higher orders.
The future cone and the antipodally identified past cone contain the same
source and differ only in the orientation of the outer affine integration.
Let $\dd\Gamma_n$ denote the complete momentum and angular measure at order
$n$, and let $D_n$ be the denominator generated by the outer affine
integration. Then
\begin{align}
\widehat Z_{n,\mathrm{cone}}^{+}
&=
\int\dd\Gamma_n\,\widehat{\mathcal J}_n^{\mathrm{YF}}
\frac{\ii}{D_n+\ii0},
\label{eq:future_cone_common_kernel}
\\
\widehat Z_{n,\mathrm{cone}}^{-}
&=
\int\dd\Gamma_n\,\widehat{\mathcal J}_n^{\mathrm{YF}}
\frac{\ii}{D_n-\ii0}.
\label{eq:past_cone_common_kernel}
\end{align}
Their sum is
\begin{equation}
\boxed{
\widehat Z_{n,\mathrm{cone}}^{\mathrm{YF}}
\equiv
\widehat Z_{n,\mathrm{cone}}^{+}
+
\widehat Z_{n,\mathrm{cone}}^{-}
=
2\ii\int\dd\Gamma_n\,\widehat{\mathcal J}_n^{\mathrm{YF}}
\PV\frac{1}{D_n}.
}
\label{eq:PV_condition_III}
\end{equation}
The principal-value denominator is therefore produced by the sum of the two
opposite affine prescriptions; it is not assigned independently to either
cone.

\subsection{Order-by-order matching and recursive determination of
\texorpdfstring{$\delta a_n$}{delta a n}}
\label{subsec:order_matching_III}
The complete quantum matching equation is
\begin{equation}
\boxed{
\Delta\widehat Z_\sigma
+
\Delta\widehat Z_\Sigma
+
\widehat Z_{\rm cone}^{\rm YF}
=0,
}
\label{eq:exact_cut_cone_matching_III}
\end{equation}
or equivalently
$\Delta\widehat Z_{\rm cut}+\widehat Z_{\rm cone}^{\rm YF}=0$,
with $\Delta\widehat Z_{\rm cut}$ defined in
Eq.~\eqref{eq:complete_quantum_cut_difference_III}. The flux contribution is
therefore part of the cut side of the equation and is not a cone source.

Write
\begin{equation}
\widehat{\mathcal Z}^{-}
=
\sum_{j\geq1}\ep^j\widehat Z_j,
\qquad
\delta a_\lambda
=
\sum_{m\geq2}\ep^m
\delta a_{m,\lambda}^{\rm out}.
\label{eq:delta_a_general_series_III}
\end{equation}
For a finite perturbative truncation, define
\begin{equation}
\widehat Z_{[N]}
\equiv
\sum_{j=1}^{N}\ep^j\widehat Z_j,
\qquad
\widehat U_{\omega,[N]}
\equiv
\exp[-\ii\omega\widehat Z_{[N]}].
\label{eq:truncated_quantum_cut_III}
\end{equation}
Each $\widehat Z_{[N]}$ is self-adjoint on the chosen smeared domain, so
$\widehat U_{\omega,[N]}$ is unitary. The update from order $N-1$ to order
$N$ is represented by
\begin{equation}
\boxed{
\widehat U_{\omega,N}
\equiv
\widehat U_{\omega,[N-1]}^\dagger
\widehat U_{\omega,[N]},
\qquad
\widehat U_{\omega,[N]}
=
\widehat U_{\omega,[N-1]}
\widehat U_{\omega,N}.
}
\label{eq:incremental_unitary_factor_III}
\end{equation}
No commutativity assumption among the cut coefficients is required. Expand
\begin{equation}
\widehat U_\omega^{-}
=
\sum_{r=0}^{\infty}\ep^r
\widehat{\mathcal U}_{\omega,r},
\qquad
\widehat{\mathcal U}_{\omega,0}=\mathbf 1.
\label{eq:cumulative_unitary_expansion_III}
\end{equation}
The first coefficients are
\begin{align}
\widehat{\mathcal U}_{\omega,1}
&=-\ii\omega\widehat Z_1,
\label{eq:unitary_coefficient_1_III}
\\
\widehat{\mathcal U}_{\omega,2}
&=-\ii\omega\widehat Z_2
-\frac{\omega^2}{2}\widehat Z_1^2.
\label{eq:unitary_coefficient_2_III}
\end{align}
At order $r$, $\widehat{\mathcal U}_{\omega,r}$ depends only on
$\widehat Z_1,\ldots,\widehat Z_r$.

We first define the radiative coefficient
\begin{equation}
\boxed{
\widehat C_{n,\lambda}^{\sigma}
\equiv
[\ep^n]
\left(
\delta a_\lambda\widehat U_\omega^{-}
\right)
=
\delta a_{n,\lambda}^{\rm out}
+
\sum_{m=2}^{n-1}
\delta a_{m,\lambda}^{\rm out}
\widehat{\mathcal U}_{\omega,n-m},
\qquad n\geq2.
}
\label{eq:Cn_sigma_generating_formula}
\end{equation}
The order-$n$ flux coefficient is
\begin{equation}
\widehat Z_{n,\Sigma}^{\rm YF}
\equiv
[\ep^n]\,\Delta\widehat Z_\Sigma.
\label{eq:Zn_Sigma_definition_III}
\end{equation}
It is a known functional of incoming operators and lower-order matched
fields. Let $[\cdot]_{F,\lambda}$ denote the spatial Fourier transform
followed by the radiative helicity projection. The complete cut coefficient
is defined by
\begin{equation}
\boxed{
\widehat C_{n,\lambda}^{\rm cut}(\mathbf K)
\equiv
\left[
\widehat C_{n,\lambda}^{\sigma}
\right]_F(\mathbf K)
+
\left[
\widehat Z_{n,\Sigma}^{\rm YF}
\right]_{F,\lambda}(\mathbf K).
}
\label{eq:Cn_cut_generating_formula}
\end{equation}
Thus the scalar flux belongs to the same side as the radiative cut term.
Equation~\eqref{eq:exact_cut_cone_matching_III} gives
\begin{equation}
\boxed{
\widehat C_{n,\lambda}^{\rm cut}(\mathbf K)
=
-
\left[
\widehat Z_{n,\rm cone}^{\rm YF}
\right]_{F,\lambda}(\mathbf K).
}
\label{eq:delta_an_matching_III}
\end{equation}
Products in the $x$ representation become momentum convolutions. Solving for
the new outgoing operator yields
\begin{equation}
\boxed{
\begin{aligned}
\delta a_{n,\lambda}^{\rm out}(\mathbf K)
={}&-
\left[
\widehat Z_{n,\rm cone}^{\rm YF}
\right]_{F,\lambda}(\mathbf K)
-
\left[
\widehat Z_{n,\Sigma}^{\rm YF}
\right]_{F,\lambda}(\mathbf K)
\\
&-
\sum_{m=2}^{n-1}
\int\frac{\dd^3q}{(2\pi)^3}\,
\delta a_{m,\lambda}^{\rm out}(\mathbf q)
\left[
\widehat{\mathcal U}_{\omega_q,n-m}
\right]_F(\mathbf K-\mathbf q),
\end{aligned}
}
\label{eq:delta_an_momentum_recursion_III}
\end{equation}
where $\omega_q=|\mathbf q|$. Every quantity on the right-hand side is fixed
by the incoming free operators and lower-order matching equations.

\subsection{Checks at orders two, three, and four}
\label{subsec:checks_234}
At second order, Eq.~\eqref{eq:delta_Sigma2_zero_III} implies
\begin{equation}
\widehat C_{2,\lambda}^{\rm cut}
=
\delta a_{2,\lambda}^{\rm out},
\label{eq:C2_check_III}
\end{equation}
and hence
\begin{equation}
\boxed{
\delta a_{2,\lambda}^{\rm out}
=
-
\left[
\widehat Z_{2,\rm cone}^{\rm YF}
\right]_{F,\lambda}.
}
\label{eq:delta_a2_direct_III}
\end{equation}
At this order the Yang--Feldman prescription is automatic because the cone
source depends only on the free incoming field.

At third order,
\begin{equation}
\boxed{
\widehat C_{3,\lambda}^{\rm cut}
=
\left[
\delta a_{3,\lambda}^{\rm out}
-\ii\omega\delta a_{2,\lambda}^{\rm out}\widehat Z_1
\right]_F
+
\left[
\widehat Z_{3,\Sigma}^{\rm YF}
\right]_{F,\lambda}.
}
\label{eq:C3_check_III}
\end{equation}
Therefore,
\begin{equation}
\boxed{
\left[
\delta a_{3,\lambda}^{\rm out}
-\ii\omega\delta a_{2,\lambda}^{\rm out}\widehat Z_1
\right]_F
=
-
\left[
\widehat Z_{3,\rm cone}^{\rm YF}
+
\widehat Z_{3,\Sigma}^{\rm YF}
\right]_{F,\lambda}.
}
\label{eq:delta_a3_direct_III}
\end{equation}
The recursive radiative term is generated by the single exact product
$\delta a_\lambda\widehat U_\omega^{-}$, while the flux contribution is a
separate cut term determined from lower orders.

At fourth order,
\begin{equation}
\boxed{
\begin{aligned}
\widehat C_{4,\lambda}^{\rm cut}
={}&
\left[
\delta a_{4,\lambda}^{\rm out}
-\ii\omega\delta a_{3,\lambda}^{\rm out}\widehat Z_1
-\ii\omega\delta a_{2,\lambda}^{\rm out}\widehat Z_2
-\frac{\omega^2}{2}
\delta a_{2,\lambda}^{\rm out}\widehat Z_1^2
\right]_F
\\
&+
\left[
\widehat Z_{4,\Sigma}^{\rm YF}
\right]_{F,\lambda}.
\end{aligned}
}
\label{eq:C4_cut_explicit_III}
\end{equation}
The fourth-order matching equation follows as Eq.~\eqref{eq:order4_matching_from_exact_cut_III} in the main text,
where products are understood in the $x$ representation, or as the
corresponding convolutions after the spatial Fourier transform. The only new
unknown at this order is $\delta a_{4,\lambda}^{\rm out}$; the recursive
radiative contribution, the scalar flux term, and the cone source are all
fixed by the incoming data and lower-order equations.
%=====================================================================
%=====================================================================

\section{Scalar Yang--Feldman analogy}
\label{app:scalar_yang_feldman}
%=====================================================================

Consider a nonlinear scalar field satisfying
\begin{equation}
    \Box\phi(x)=\ep J[\phi](x).
    \label{eq:scalar_wave_YF_III}
\end{equation}
The exact field may be represented either with retarded incoming data or with
advanced outgoing data,
\begin{align}
    \phi(x)
    &=
    \phi_{\mathrm{in}}(x)
    +
    \ep\,G_{\mathrm{ret}}*J[\phi](x),
    \label{eq:scalar_ret_exact_III}
    \\
    \phi(x)
    &=
    \phi_{\mathrm{out}}(x)
    +
    \ep\,G_{\mathrm{adv}}*J[\phi](x).
    \label{eq:scalar_adv_exact_III}
\end{align}
The Yang--Feldman prescription begins by selecting the interacting field as
the retarded functional of the free incoming field:
\begin{equation}
    \boxed{
    \phi_{\mathrm{YF}}[\phi_{\mathrm{in}}]
    =
    \phi_{\mathrm{in}}
    +
    \ep\,G_{\mathrm{ret}}*
    J\!\left[\phi_{\mathrm{YF}}[\phi_{\mathrm{in}}]\right].
    }
    \label{eq:scalar_YF_retarded_field_III}
\end{equation}
This equation is solved recursively. Once the retarded field has been
selected, the advanced representation does not introduce a different field
inside the source. It uses the same functional
$\phi_{\mathrm{YF}}[\phi_{\mathrm{in}}]$:
\begin{equation}
    \phi_{\mathrm{YF}}[\phi_{\mathrm{in}}]
    =
    \phi_{\mathrm{out}}
    +
    \ep\,G_{\mathrm{adv}}*
    J\!\left[\phi_{\mathrm{YF}}[\phi_{\mathrm{in}}]\right].
    \label{eq:scalar_YF_advanced_same_source_III}
\end{equation}
Subtracting the two representations gives
\begin{equation}
    \boxed{
    \phi_{\mathrm{out}}-\phi_{\mathrm{in}}
    =
    \ep\left(G_{\mathrm{ret}}-G_{\mathrm{adv}}\right)*
    J\!\left[\phi_{\mathrm{YF}}[\phi_{\mathrm{in}}]\right].
    }
    \label{eq:scalar_YF_scattering_III}
\end{equation}
The source is evaluated on the same retarded incoming-data field in both
representations; only the outer Green function is changed.

Perturbatively, writing $\phi = \sum_{n=0}^\infty \ep^n \phi_n$ with
$\phi_0 = \phi_{\mathrm{in}}$, Eq.~\eqref{eq:scalar_YF_retarded_field_III}
determines $\phi_{\mathrm{YF}}[\phi_{\mathrm{in}}]$ order by order. Substitution
in Eq.~\eqref{eq:scalar_YF_scattering_III} then gives the scattering operator
entirely in terms of incoming asymptotic fields, without an independent
advanced-field reconstruction.

%=====================================================================
\section{Shear normalization and null-cone News modes}
\label{app:shear_null_cone_normalization}
We spell out the frequency dependence connecting the shear convention
\eqref{eq:main_shear_cartesian_normalization} with the News algebra. For
$\omega>0$, the Cartesian delta distribution satisfies
\begin{equation}
\delta^{(3)}(\mathbf k-\mathbf k')
=\frac{1}{\omega^2}\delta(\omega-\omega')
\delta_{S^2}^{(2)}(\Omega,\Omega').
\end{equation}
Therefore the frequency--angle oscillators obey
\begin{equation}
[a_\lambda(\omega,\Omega),a^\dagger_{\lambda'}(\omega',\Omega')]
=\frac{\delta_{\lambda\lambda'}}{\omega}
\delta(\omega-\omega')\delta_{S^2}^{(2)}(\Omega,\Omega').
\end{equation}
A shear expansion with an explicit $1/\sqrt{\omega}$ is consistent
with delta-normalized frequency--angle modes. To see the conversion,
define $c_\lambda(\omega,\Omega)=\sqrt{\omega}\,a_\lambda(\omega,\Omega)$.
Then
\begin{equation}
[c_\lambda(\omega,\Omega),c^\dagger_{\lambda'}(\omega',\Omega')]
=\delta_{\lambda\lambda'}\delta(\omega-\omega')
\delta_{S^2}^{(2)}(\Omega,\Omega'),
\qquad \frac{c_\lambda(\omega,\Omega)}{\sqrt{\omega}}
=a_\lambda(\omega,\Omega).
\end{equation}
Thus the square-root factor can be used with the delta-normalized
frequency--angle modes, while the Cartesian-momentum CCR used here requires
its absorption into the mode definition. Retaining $1/\sqrt{\omega}$ in
the shear expansion while identifying the mode with the unrescaled
Cartesian operator would combine different conventions and alter the shear
commutator. This is a convention conversion; the tensor-to-dyad and Fourier
normalizations needed for an exact coefficient comparison with Ashtekar's
formulation must also be specified.

With $\bar\sigma=\sigma^\dagger$ and the two helicities in
Eq.~\eqref{eq:main_shear_cartesian_normalization}, this gives, as a
smeared distribution,
\begin{align}
[\sigma(u,\Omega),\bar\sigma(u',\Omega')]
&=-\frac{2iG}{\pi}\delta_{S^2}^{(2)}(\Omega,\Omega')
\int_0^\infty\frac{d\omega}{\omega}\sin[\omega(u-u')]\\
&=-iG\,\operatorname{sgn}(u-u')\,
\delta_{S^2}^{(2)}(\Omega,\Omega').
\label{eq:shear_sign_normalization}
\end{align}
The result uses the unit-sphere area measure, $\hbar=1$, and the displayed
shear normalization. It holds in the nonzero-frequency radiative sector;
zero-mode contributions require separate boundary data.

For comparison, Prabhu and Satishchandran use null-cone coordinates
$p^a=\omega(1,\widehat{\mathbf p})$ and News modes whose commutator is
proportional to $\omega\delta_{\mathcal C^+}(p,p')$ times the symmetric
trace-free projector \cite{PrabhuSatishchandran2024}. Their cone delta is
normalized with respect to $d\omega\,d\Omega$, so
\begin{equation}
\delta_{\mathcal C^+}(p,p')
=\omega^2\delta^{(3)}(\mathbf p-\mathbf p').
\end{equation}
Their News-mode commutator therefore scales as
$\omega^3\delta^{(3)}(\mathbf p-\mathbf p')$. In our convention a
positive-frequency News mode is proportional to $\omega a_\lambda$;
its commutator has the same Cartesian frequency dependence. The exact
coefficient requires translating the tensor shear, polarization dyad,
Fourier normalization, and the factor in the definition of News; it is not
fixed by this scaling comparison alone. The background-null-vector
normalization $\ell^0=1/\sqrt2$ used in BKR changes the relation between
its Fourier parameter and $|\mathbf k|$ by a constant factor.

\end{document}